\documentclass[%
 amsmath,amssymb,
 aps,
 showpacs, showkeys, nofootinbib, longbibliography
]{revtex4-1}

\usepackage{url}
\usepackage{graphicx}
\usepackage{bm}
\usepackage{hyperref}
\usepackage{tabularx}

\begin{document}

\title{The Completeness of Quantum Mechanics and the Determinateness and Consistency of Intersubjective Experience: Wigner's Friend and Delayed Choice}

\author{Michael Silberstein}
\email{silbermd@etown.edu}
\affiliation{
 Department of Philosophy\\ Elizabethtown College\\Elizabethtown, PA 17022\\
}%
\affiliation{
Department of Philosophy\\ University of Maryland\\ College Park, MD 20742\\
}%
\author{W.M. Stuckey}
\affiliation{
 Department of Physics\\ Elizabethtown College\\Elizabethtown, PA 17022\\
}%

\date{\today}

\begin{abstract}
 Recent experiments (gedanken or otherwise) and theorems in quantum mechanics (QM), such as new iterations on Wigner's friend and delayed choice, have led many people to claim that QM is not compatible with determinate and intersubjectively consistent experience (what some call absoluteness of observed events), such as experiences of experimental outcomes. In the case of delayed choice the tension is between our experience of free will and a possible ``superdeterminism'' at work in QM. At the very least, some have suggested that the only way to save absoluteness of observed events, is to give up one or more of the following assumptions: free will, locality, or the completeness of QM. Our goal in this paper is to provide a take on QM that explains why there is and must always be determinate and intersubjectively consistent experience about all experimental outcomes (absoluteness of observed events). Our take accepts the completeness of the theory and requires no invocation of relative states (e.g., outcomes being relative to branches, conscious observers, etc.). And finally, this take requires no allegedly hybrid models such as claims about ``subjective collapse.'' We provide a take on QM that yields a single world wherein all the observers (conscious or otherwise) agree about determinate and definite outcomes, because those outcomes are in fact determinate and definite. We provide a realist psi-epistemic take on QM that saves the absoluteness of observed events and the completeness of QM, without giving up free will or locality. We also show how our realist psi-epistemic account eliminates the measurement problem and, coupled with our take on neutral monism, also eliminates the hard problem of consciousness. The key to all this is to let go of the following offending assumptions: 1) physicalism, 2) fundamentalism, and relatedly 3) dualism about conscious experience, 4) the notion that fundamental explanation is always constructive, causal or dynamical, and relatedly, 5) realism about the wavefunction. Together these assumptions force us into the hard problem, they force us into the measurement problem, and they force us to seek the solutions to these problems in fundamental physics, e.g., by trying to relate these problems to one another directly, with very little success. Sometimes, when a problem is deeply intractable the best move is to jettison the offending assumptions that led to the problem in the first place. This is precisely what we do herein.     
\end{abstract}

\keywords{Wigner's friend, delayed choice quantum eraser, measurement problem, hard problem of consciousness, neutral monism, locality, superdeterminism, free will, completeness of QM, realist psi-epistemic, realism, principle versus constructive explanation, spacetime}
\maketitle

\section{\label{intro1}Introduction}
For a variety of reasons, there is renewed interest in trying to relate conscious experience to quantum mechanics (QM). Usually someone tries to explain one in terms of the other, e.g., they claim that conscious experience collapses the wavefunction \cite{mcqueen}. There are also many attempts to explain conscious experience or some feature of it in terms of some quantum process or property such as wavefunction collapse, entanglement, etc. Such accounts are metaphysically diverse, ranging from strong emergence, to panpsychism, to dual-aspect theories and beyond \cite{atmanspacher}. Obviously, attempts at explaining some feature of QM in terms of some feature of conscious experience or vice-versa, require that one be very specific about the interpretation of QM at issue and the particular feature/conception of conscious experience in question. Often when this game is played one assumes that wavefunction realism is true and that qualia is the best way to think about conscious experience; we will jettison both those assumptions in this paper. There are of course other ways to relate QM and conscious experience, such as simply using the Hamiltonian formalism of QM to model conscious decision making, e.g., thinking thru a decision is like being in a superposition state and making a decision is like wavefunction collapse \cite{atmanspacher}. 

More generally, Gao argues that the very measurement problem itself should be characterized as a problem about determinate-experience \cite[p. 4]{gao}:
\begin{quote}
[t]he problem is not only to explain how the linear dynamics can be compatible with the appearance of definite measurement results obtained by physical devices, but also, and more importantly, to explain how the linear dynamics can be compatible with the existence of definite experiences of conscious observers.
\end{quote}
Of course one could certainly push back on Gao's characterization of the measurement problem. Many would argue that the measurement problem would exist even in a world with no conscious observers or that there are interpretations of QM that yield definite classical outcomes and thus determinate experiences. However, if one assumes wavefunction realism as Gao does and one adopts some no collapse interpretation of QM, then it is certainly reasonable to raise the question of why and how we have determinate experiences. Even though one may be convinced in general that no-collapse interpretations such as Many-Worlds and Bohmian mechanics have no such concern, it is this question of determinate experience and intersubjectively consistent determinate experience that we shall take up herein. This is because, as we are all well aware, there are cases (experiments, gedanken or otherwise) where, at least on certain interpretations, there is a problem about determinate and intersubjectively consistent experience, such as the appearance of definite measurement outcomes that all observers can experience as determinate and will agree upon. Herein we will look at two such cases: Wigner's friend and delayed choice quantum eraser with a twist (see also Lucien Hardy's chapter in this volume for more on the latter). We will discuss why these cases raise concerns for the determinateness and intersubjective consistency of conscious experience at length later. Let us note for now that Wigner's friend has recently raised its ugly head again. There are new Wigner's friend related experiments, alleging to show that QM is either incomplete or incompatible with determinate and intersubjectively consistent conscious experience, such as experiences of measurement outcomes \cite{proietti,Bong2020}. Recently Hardy has suggested putting conscious decision makers into an EPR delayed-choice type experiment to see if they could violate the predictions of QM \cite{Hardy2017}. Such an experiment pits the completeness of QM against our everyday experience of free will. Hardy discusses this possibility further in this volume and we will consider a much more idealized version of such an experiment herein.  

The specific question we want to address is as follows: is there a take on QM that explains why there is and must always be determinate and intersubjectively consistent experience about all experimental outcomes? A take that accepts the completeness of the theory (i.e., no modification to the formalism is made such as the addition of an objective collapse mechanism), and furthermore requires no invocation of relative states (e.g., outcomes being relative to branches, conscious observers, etc.). And finally, a take that requires no allegedly hybrid models such as claims about ``subjective collapse.'' We want a take on QM that provides a single world wherein all the observers (conscious or otherwise) agree about determinate and definite outcomes, because those outcomes are in fact determinate and definite. We also want a take on QM that explains why conscious human choices cannot violate the predictions of QM, that is nonetheless consistent with our experience of having free will. Herein, we will provide such a take on QM. Our take on QM is a realist psi-epistemic one that is based on principle explanation as opposed to constructive, dynamical or causal mechanical explanation. As we shall see, the principles in question constitute adynamical global constraints ranging over spacetime, not unlike those constraints underlying special relativity (SR). The idea of a principle-based take on QM is not new to us, the quantum information theory (QIT) community has sought such an account for a long time \cite{fuchs1}. Our principles constrain what observers (conscious or otherwise) can experience, measure, observe, etc. One such constraint is that no observer occupies a preferred frame of reference. Our take on QM has many additional advantages such as a deflation of the measurement problem, a local explanation of EPR-Bell correlations in terms of adynamical global constraints, and an explanation of the Born rule (for details see \cite{ourbook, NMEntropy2020, MerminChallenge}). The key to getting all these advantages is to give up naive realism about the wavefunction, which is motivated in part by physicalism and the little questioned assumption that fundamental explanation must always be constructive, causal or dynamical. Let us note again that while our take on QM is psi-epistemic, unlike QBism or pragmatic accounts, our take is fully realistic and resides fully in spacetime.   
     
There is also another advantage of our particular principle take on QM and relativity as well. We are now in a position to jettison the second odious assumption that forces an eternal return of the dialogue about how consciousness might explain the collapse of the wavefunction or vice-versa. The first offending assumption as we noted, is realism about the wavefunction coupled with collapse. The second troublemaking assumption is dualism about conscious experience, i.e., the idea that conscious experience is best conceived as some sort of qualia that is either somehow produced by brains via strong emergence or must be present in some form in fundamental physics a la panpsychism. Both strong emergence and panpsychism are notoriously problematic and yet, for want of a better alternative, we return to them over and over again \cite{NMEntropy2020}. Our view provides a better alternative called neutral monism. The basic idea of neutral monism is that the mental and physical are nondual (not essentially different), they are neutral in virtue of being neither essentially mental nor physical, and in virtue of the fact that they are grounded in something neutral \cite{NMEntropy2020}. This ubiquitous``something'' does not ``pervade'' spacetime, it does not ``generate'' spacetime, it is co-extensional with spacetime. According to neutral monism, our conscious experience of an external world is not some virtual model or construction of the world trapped in the mind and generated by the brain, while the actual external physical world lies outside us forever, as I-know-not-what noumena. Neutral monism is a form of direct realism, wherein there is only the spatiotemporally extended world of experience, of which we are inextricably a part; the discipline of physics is all about the world so described. 

It will become clear as we go along what all this entails about the relationship between physics and psychology. For now however, note that given neutral monism, physics doesn’t start with experience merely for pragmatic reasons. Physics is inherently all about the possibility of and rules of experience, but not because the world is mind dependent. Just as there is no metaphysical dualism of the ``inner'' world of experience and the ``outer'' physical world, there is no inherent dualism of psychology and physics. As Bertrand Russell puts it, ``The whole duality of mind and matter ... is a mistake; there is only one kind of stuff out of which the world is made, and this stuff is called mental in one arrangement, physical in the other'' \cite[p. 15]{taylor1996}. Compare this to the words of William James, ``Things and thoughts are not fundamentally heterogeneous; they are made of one and the same stuff, stuff which cannot be defined as such but only experienced; and which one can call, if one wishes, the stuff of experience in general. ... `Subjects' knowing `things' known are `roles' played, not `ontological' facts'' \cite[p. 63]{carnap}. The point is that when Galileo and others insisted on the primary/secondary quality distinction or any other kind of dualism of inner mental experience and external physical world, they doomed us to the mind/body problem and the hard problem. Such dualism is not empirical data we experience directly, it is a cognitive illusion, an inductive projection of theorizing minds. The way to undo this mistake is not to merely move the location of the mysterious dualism from brains (strong emergence) to fundamental physics (panpsychism), but to reject completely and thoroughly the primary/secondary property distinction altogether with neutral monism. Given neutral monism, what we call physical phenomena are just one mode of description of the neutral base. 

We hope the reader will see that if one is willing to give up the assumptions of wavefunction realism and dualism about conscious experience, the very assumptions that have led us into this morass that includes the hard problem, the mysteries of QM and recurring idea that they are somehow deeply related, there is a way out to an entirely different picture of the world and our place in it. Given neutral monism, conscious experience and physics are in fact deeply related, but not in the occult way most people think. Rather, they are deeply related because conscious experience and the physical world are nondual, which means that physics is inherently already all about experience and constraints upon it. It is not hard to see that one of the things motivating these problematic assumptions is physicalism (the idea that everything is physical or at least that physical phenomena are basic) and fundamentalism (the idea that all axiomatic facts must reside in basic physics). Given these assumptions, it isn't surprising that we continue to chase our tail seeking deep connections between the hard problem and the measurement problem, that we keep gravitating toward panpsychism, etc. Neutral monism is a hard rejection of both these assumptions. Sometimes, when a problem remains intractable over long periods of time, the best move is to deflate the problem by rejecting the axiomatic assumptions that got you there. This paper is an attempt to deflate both the hard problem and the measurement problem in one fell swoop.     

In section \ref{BongEtAl} we will use a recent, much discussed Wigner's friend article that alleges to show we must either give up the completeness of QM, commonsense intuitions about free will, locality, or the determinate and objective nature of reality. This is a very useful set-up for our paper since we allege to save all these assumptions, with the possible exception of commonsense intuitions about free will--the reader will decide. In section \ref{intro2} we will introduce the reader to the concept of principle explanation and briefly explain our particular local, principle explanation of EPR-Bell correlations. This will allow the reader to appreciate our analysis of various Wigner's friends experiments and the quantum eraser delayed choice experiment with a twist, both to be discussed in subsequent sections. In section \ref{Conclusion} we will summarize and recontextualize everything that we have said in light of neutral monism, in order to show the reader what a different and better world it is.


\section{\label{BongEtAl}Wigner's Friend Revisited}

Let us recall the Wigner's friend thought experiment. We start with the Schr{\"o}dinger’s cat thought experiment (for example) and now we imagine a human observer comes up to the box and opens the lid. If we assume that QM is complete and thus applies to all macroscopic measuring devices including conscious observers, then the Schr{\"o}dinger equation tells us we will now have an entangled state that includes the human observer and the cat in a box, and so on if we keep adding observers. This is of course just the measurement problem. While the Wigner's friend thought experiment specifically raises the question as to whether a conscious observer could ever be in a QM superposition or entangled state, note that the measurement problem as such does not essentially involve consciousness in any way, e.g., we could have a sent a robot to look in the box and made the same point. However, some people have speculated that if conscious states cannot be placed in a superposition state, perhaps that is evidence that consciousness itself collapses the wavefunction. This is the question Bong et al. hope to explore in a new experiment, the results of which were recently published in Nature \cite{Bong2020}. The experiment is a toy one in the sense that in the place of conscious observers or macroscopic measuring devices, they use photons. They believe however to have a ``proof-of-principle'' that hopefully someday can be scaled up to macroscopic measuring devices and conscious observers. What is it they allege to have proved?

Bong et al. allege to show that we must give up at least one of the following assumptions:
\begin{itemize}
\item Assumption 1 (Absoluteness of Observed Events (AOE)): An observed event is a real single event, and not relative to anything or anyone.  

\item Assumption 2 (No-Superdeterminism (NSD)): Any set of events on a space-like hypersurface is uncorrelated with any set of freely chosen actions subsequent to that space-like hypersurface.

\item Assumption 3 (Locality (L)): The probability of an observable event \textit{e} is unchanged by conditioning on a space-like-separated free choice \textit{z}, even if it is already conditioned on other events not in the future light-cone of \textit{z}.

\item Assumption 4 (The completeness of QM): QM unmodified applies to any and all macroscopic measuring devices including human observers. 
\end{itemize}
We will talk about the details of this experiment in section \ref{WFexperiment}, but for now we merely want to use it to frame our paper. In order to appreciate why some people might think this result is a big deal, we first need to remind ourselves what the Kochen-Spekker (KS) and Bell theorems are supposed to show. KS is supposed to show that measurement outcomes (possessed values) in QM must be contextual. That is, KS is generally taken to show that measurement outcomes (possessed values) in QM are never independent of the measurement context--how the value is eventually measured. KS then is generally taken to be a knock on a hidden variables account of QM. In addition to that, Bell's theorem is supposed to establish that even if one is prepared to concede contextuality, any such explanation must be non-local, i.e., there can be no local hidden variable account for QM. It is well known that both proofs have loopholes and here's the point. Both proofs take as tacit truths some or all of assumptions 1-4 above. The point being that you can get around one or another of the two theorems if you are willing to give up one or more of assumptions 1-4. But obviously there is a price, for each assumption has its strong supporters. 

Here is why you might not think this result is a big deal: we already knew all this. Indeed, most every existing interpretation of QM is defined by giving up one or more of these four assumptions. Defenders of the result's importance will note that the result is based on weaker constraints than those assumed by either KS or Bell. What Bong et al. allege to have shown is that given assumption 4, given the claim that QM applies to macroscopic measuring devices and conscious observers, then we must give up one or more of assumptions 1-3. Detractors will reply that since photons are neither macroscopic nor conscious, the new experiment proves absolutely nothing new until at the very least, they scale it up. Regardless of one's opinion, we can say that the new experiment does at least put our choices in stark relief. However, no one has of yet established that this list of assumptions is exhaustive, that is, there may be other choices we could make, other things we could give up. For example, notice that built into assumption 4 is the suppressed assumption that we are realist's about QM. QBists and others reject this assumption and thus reject the idea that QM can tell us what other observers will measure, or they will say QM is only about subjective belief states, not physical states or brain states.  

Other people make stronger claims about what the new experiment shows. For example, Renato Renner claims the new theorem is telling us that QM needs to be replaced \cite{Merali2020}. We will show that such claims are patently false. QM doesn't need to be replaced, it needs to be reconceived. We will analyze Bong et al. and other related new Wigner's friend type experiments with our reconception of QM in order to show that we can keep every assumption above except possibly number 2. However, we need to make clear that it is wrong to call this assumption ``superdeterminism,'' rather it should be called ``measurement independence.'' Bell in his proof assumes that the properties of a particle are independent of the (future) measurements to be performed on them, this is measurement independence (MI). Often dubbed ``retrocausal,'' there are accounts of QM that reject this assumption in order to provide a local take on EPR-Bell correlations that is consistent with SR.  The soul of this idea is as follows \cite{PriceWharton}: 
\begin{quote}

In putting future and past on an equal footing, this kind of approach is different in spirit from (and quite possibly formally incompatible with) a more familiar style of physics: one in which the past continually generates the future, like a computer running through the steps in an algorithm. However, our usual preference for the computer-like model may simply reflect an anthropocentric bias. It is a good model for creatures like us, who acquire knowledge sequentially, past to future, and hence find it useful to update their predictions in the same way. But there is no guarantee that the principles on which the universe is constructed are of the sort that happens to be useful to creatures in our particular situation. Physics has certainly overcome such biases before—the Earth isn’t the center of the universe, our sun is just one of many, there is no preferred frame of reference. Now, perhaps there’s one further anthropocentric attitude that needs to go: the idea that the universe is as ``in the dark'' about the future as we are ourselves.
\end{quote}
While one can find many different instantiations of this idea in the literature, notice that nothing in this description entails any particular mechanism such as literal causal influences or signals coming from the future. Nor does this idea entail superdeterminism. Technically speaking, a superdeterministic world is one in which independence is violated via a past common cause—a common cause of one’s choice of measurements and say the particle spin properties, in the case of Bell correlations. In short, superdeterminism is a conspiratorial theory with only past-to-future causation. So while superdeterminism does entail that experimenters are not free to choose what to measure without being influenced by events in the distant past, i.e., it does give up MI, it does so in a particularly spooky way, forcing us to accept some very special conditions at the big bang as a brute fact or seek some sort of physically acceptable explanation for those initial conditions that is presumably not some sort of supernatural conspiracy. In the last few years we have argued that there are really two ideas buried in retrocausal accounts. One is the idea that we should seek out some dynamical or constructive explanation that involves determination relations from the future. The other idea, our idea, is that we should jettison dynamical, causal and constructive explanation altogether when it comes to EPR-Bell correlations and instead seek an explanation in terms of adynamical global constraints \cite{stuckeyIJQF}. This would be an adynamical and acausal explanation that seeks a constraint-based explanation for the pattern of EPR-Bell correlations in spacetime as given by QM. The idea here isn't merely the retrocausal idea that maybe the future determines the past as much as the other way around. The idea is to stop worrying about determination being time-like versus space-like, it can be both or neither, because the nature of explanation in the case of EPR-Bell correlations is not in any way causal or dynamical. Therefore, even the focus on measurement independence is somewhat misleading, because it immediately makes people think about future measurements somehow retrocausally bringing about the properties of QM systems in the past, when the point is to view the entire experiment in terms of spatiotemporal constraints that don't care about time-like versus space-like. 

Before we can explain how the Wigner’s friend and delayed choice quantum eraser gedanken experiments of QM bear on conscious experience per our constraint-based approach to physics, we must provide some background. Our perceptions are formed dynamically, i.e., in a causal, temporally sequential fashion, as in a game of chess. For example, move number 25 cannot be made until all moves 1 – 24 have been made. Therefore, move number 25 can be said to be explained by move number 24 which is explained by move number 23, etc. In contrast, the answers to a crossword puzzle explain each other, no word has any explanatory priority over the other words. A crossword puzzle is analogous to what we mean by ``adynamical'' or ``constraint-based'' explanation. The goal in a crossword puzzle is a self-consistent collection of intersecting words in accord with the clues (constraints) given. We propose interpreting modern physics in analogous fashion where the constraints in physics are spatiotemporal and can be understood as constraints on experience per neutral monism. Do such explanations also place a constraint on free will? Yes, of course, as we will show explicitly in our analysis of the delayed choice QM eraser experiment with a twist. But so what? Who thought physics was unquestionably compatible with libertarian free will anyway? Who believed that the laws of physics placed no constraints on the degrees of freedom of human action? However, let us note again, this is not superdeterminism, fatalism or any other kind of conspiracy theory. It is simply the acknowledgement that not all determination relations are dynamical or causal.  

We have provided such an adynamical global constraint explanation for EPR-Bell correlations, QM superposition such as seen in the twin-slit experiment, and other QM phenomena in a variety of different publications over the years \cite{ourbook,stuckeyIJQF,NMEntropy2020,stuckeyICP,MerminChallenge}. As we will illustrate in the next section, most recently we have simplified our adynamical global constraint-based explanation and think it is best appreciated as a principle explanation of the sort we find in SR.

Our principled adynamical global constraints are as follows:
\begin{itemize}
\item Axiom 1: Interacting ``bodily objects'' coextensive with space and time form the context of all self-consistent, shared perceptual information between POs (point of observational origin--need not be conscious) and these perceptions constitute different reference frames in that spacetime model of the Physical (the ``real external world'').
\item Axiom 2: For all of physics, the ``perceptions'' of any particular PO do not provide a privileged perspective of the Physical. This is known as ``no preferred reference frame'' (NPRF).
\end{itemize}
We will discuss these axioms at greater length going forward, but Axiom 1 is basically saying that what QM is really telling us is that to exist (to be a diachronic entity in space and time) is to interact with the rest of the universe creating a consistent, shared set of classical information constituting spacetime. It is in effect a generalization of ``the boundary of a boundary principle'' $\partial\partial  = 0$ \cite{misner,wise,BBP}. Axiom 2 is just a generalization of the relativity principle.  
 
Putting this all together we can say the following. We understand that each subject is just a conscious point of origin (PO) in spacetime. The ``perceptions'' of each PO form a context of interacting trans-temporal objects (TTOs) for that PO. Since TTOs are ``bodily objects'' with worldlines in spacetime, TTOs are coextensive with space and time. When POs exchange information about their ``perceptions,'' they realize that some of their disparate ``perceptions'' fit self-consistently into a single spacetime model with different reference frames for each PO. Thus, physicists’ spacetime model of the ``Physical'' represents the self-consistent collection of shared perceptual information between POs, e.g., ``perceptions'' upon which Galilean or Lorentz transformations can be performed. That is, spacetime is the self-consistent collection of shared classical information regarding diachronic entities (classical objects), which interact per QM. The consistency of shared classical information of spacetime is guaranteed by the divergence-free (gauge invariant) nature of the adynamical global constraints for classical and quantum physics. We understand that the reader will no doubt be puzzled by all this now. As we unpack and apply these ideas we believe it will all come into focus. We will begin with our principle explanation of EPR-Bell correlations in the next section.

Finally however, it is important that the reader appreciate our realistic, psi-epistemic account of QM in a little more detail \cite{stuckeyIJQF}. Recall that the measurement problem and the worry about Wigner's friend are driven by wavefunction realism. We need to make it clear what our alternative looks like. If one constructs the differential equation (Schr{\"o}dinger equation) corresponding to the Feynman path integral, the time-dependent foliation of spacetime gives the wavefunction $\Psi (x,t)$ in concert with our time-evolved perceptions and the fact that we do not know when the outcome is going to occur. Once one has an outcome, both the configuration $x_o$, that is the specific spatial locations of the experimental outcomes, and time $t_o$ of the outcomes are fixed, so the wavefunction $\Psi (x,t)$ of configuration space becomes a probability amplitude $\Psi (x_o,t_o)$ in spacetime, i.e., a probability amplitude for a specific outcome in spacetime. Again, the evolution of the wavefunction in configuration space before it becomes a probability amplitude in spacetime is governed by the Schr{\"o}dinger equation.

However, the abrupt change from wavefunction in configuration space to probability amplitude in spacetime is not governed by the Schr{\"o}dinger equation. In fact, if the Schr{\"o}dinger equation is universally valid, it would simply say that the process of measurement should entangle the measurement device with the particle being measured, leaving them both to evolve according to the Schr{\"o}dinger equation in a more complex configuration space (as in the relative-states formalism shown below). The Many-Worlds interpretation notwithstanding, we do not seem to experience such entangled existence in configuration space, which would contain all possible experimental outcomes. Instead, we experience a single experimental outcome in spacetime.

This contradiction between theory and experience is called the ``measurement problem.'' However, the time-evolved story in configuration space is not an issue with the path integral formalism as we interpret it, because we compute $\Psi (x_o,t_o)$ directly. That is, in asking about a specific outcome we must specify the future boundary conditions that already contain definite and unique outcomes. Thus, the measurement problem is a non-starter for us. When a QM interpretation assumes the wavefunction is an epistemological tool rather than an ontological entity, that interpretation is called ``psi-epistemic.'' In our path integral, constraint-based account the wavefunction in configuration space is not even used, so our account is trivially psi-epistemic. But, one must also fully understand the classical-quantum contextual implications of our view as expressed in Axiom 1 above. The so-called ``quantum system'' is in fact the totality of the entire experimental set-up \cite[p. 738]{gomatam} such that different set-ups or configurations are not probing some autonomous quantum realm, but actually constitute different ``systems.'' QM then, just as the KS theorem and more recent related theorems suggest, is a theory that has quantum-classical contextuality at its heart, that is among its deepest lessons about the nature of physical reality. As Ball puts it, ``the quantum experiment is not probing the phenomenon but is the phenomenon'' \cite[p. 90]{ball}. Given our Lagrangian approach, the entire experimental set-up includes future boundary conditions: the experiment from initiation to termination in spacetime, no time-evolved configuration space required. For a detailed treatment of the measurement problem, the Born rule, and environmental decoherence see \cite{ourbook}.

Finally, we should clarify possible sources of confusion with this quantum-classical contextuality. First, we are not saying that there exist quantum entities with inaccessible properties that are traversing the space between emitters and receivers. We are denying the existence of any worldlines for quantum momentum exchanges, no matter how large the energy or momentum being exchanged. On pain of infinite regress, a quantum exchange of momentum is not a TTO. A quantum of momentum or energy does not have any ``intrinsic properties'' or intrinsic existence as we discussed above as regards Axiom 1. Second, there is no Bohrian quantum-classical ``cut'' and they do not exist independently of each other. As we are all aware, there doesn't seem to be any limit on the size of the quantum momentum exchange.  

Ironically perhaps, the adynamical global constraints in question actually guarantee a sensible world of classical/dynamical objects that evolve in space and time. That is, classical/dynamical objects (things that persist), time, and space are all interdependent just as general relativity suggests. In short, we will explain why, on our view, this is and must be a world with determinate experience and universal intersubjective agreement.


\section{\label{intro2}Principle Explanation of EPR-Bell Correlations}

Many physicists in quantum information theory (QIT) are calling for ``clear physical principles'' \cite{fuchs1} to account for QM. As Hardy points out, ``The standard axioms of [quantum theory] are rather ad hoc. Where does this structure come from?''\cite{hardy2016} Fuchs points to the postulates of SR as an example of what QIT seeks for QM \cite{fuchs1} and SR is a principle theory \cite{felline}. That is, the postulates of SR are constraints offered without a corresponding constructive explanation. In what follows, Einstein explains the difference between the two \cite{einstein1919}:
\begin{quote}
We can distinguish various kinds of theories in physics. Most of them are constructive. They attempt to build up a picture of the more complex phenomena out of the materials of a relatively simple formal scheme from which they start out. ...\\

Along with this most important class of theories there exists a second, which I will call ``principle-theories.'' These employ the analytic, not the synthetic, method. The elements which form their basis and starting point are not hypothetically constructed but empirically discovered ones, general characteristics of natural processes, principles that give rise to mathematically formulated criteria which the separate processes or the theoretical representations of them have to satisfy. ...\\

The advantages of the constructive theory are completeness, adaptability, and clearness, those of the principle theory are logical perfection and security of the foundations. The theory of relativity belongs to the latter class.
\end{quote}
For those who believe the fundamental explanation for QM phenomena must be constructive, at least in the sense envisioned by Einstein above, none of the mainstream interpretations neatly fit the bill. Not only do most interpretations entail some form of QM holism, contextuality, and/or non-locality, the remainder invoke priority monism and/or multiple branches or outcomes. The problem with attempting a constructive account of QM is, as articulated by Van Camp, ``Constructive interpretations are attempted, but they are not unequivocally constructive in any traditional sense'' \cite{vancamp2011}. Thus, he states \cite{vancamp2011}: 
\begin{quote}
    The interpretive work that must be done is less in coming up with a constructive theory and thereby explaining puzzling quantum phenomena, but more in explaining why the interpretation counts as explanatory at all given that it must sacrifice some key aspect of the traditional understanding of causal-mechanical explanation.
\end{quote}
It seems clear all of this would be anathema to Einstein and odious with respect to constructive explanation, especially if say, statistical mechanics is the paradigm example of constructive explanation. Thus, for many it seems wise to at least attempt a principle explanation of QM, as sought by QIT. The problem with QIT's attempts is noted by Van Camp \cite{vancamp2011}:
\begin{quote}
However, nothing additional has been shown to be incorporated into an information-theoretic reformulation of QM beyond what is contained in QM itself. It is hard to see how it could offer more unification of the phenomena than QM already does since they are equivalent, and so it is not offering any explanatory value on this front.
\end{quote}

The term ``reference frame'' has many meanings in physics related to microscopic and macroscopic phenomena, Galilean versus Lorentz transformations, relatively moving observers, etc. The difference between Galilean and Lorentz transformations resides in the fact that the speed of light is finite, so NPRF entails the light postulate of SR, i.e., that everyone measure the same speed of light \textit{c}, regardless of their motion relative to the source. If there was only one reference frame for a source in which the speed of light equaled the prediction from Maxwell's equations ($c = \frac{1}{\sqrt{\mu_o\epsilon_o}}$), then that would certainly constitute a preferred reference frame. Herein, we extend NPRF to include the measurement of another fundamental constant of nature, Planck's constant \textit{h} ($=2\pi\hbar$). 

\begin{figure}
\begin{center}
\includegraphics [height = 55mm]{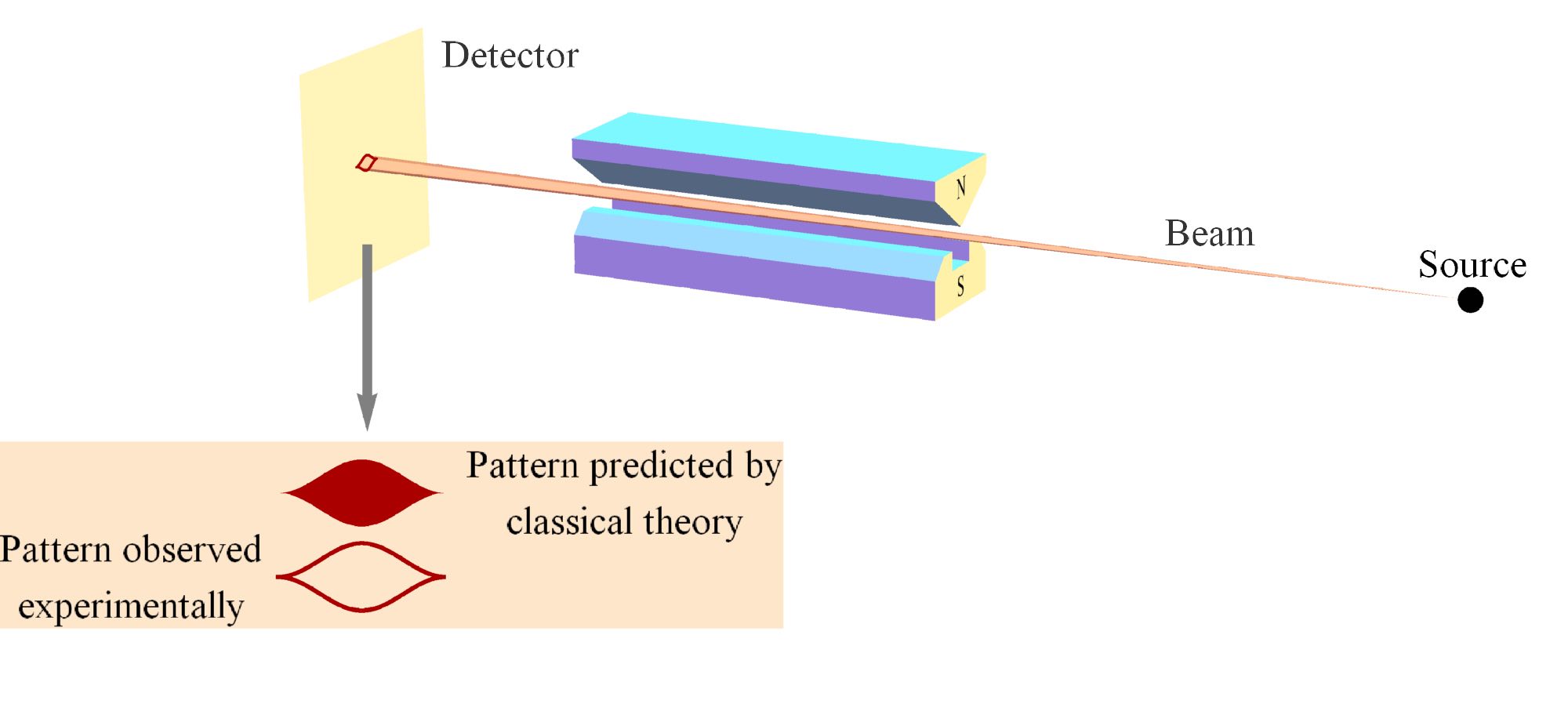}  \caption{A Stern-Gerlach (SG) spin measurement showing the two possible outcomes, up ($+\frac{\hbar}{2}$) and down ($-\frac{\hbar}{2}$) or $+1$ and $-1$, for short. The important point to note here is that the classical analysis predicts all possible deflections, not just the two that are observed. This binary (quantum) outcome reflects Dakic and Brukner's first axiom in their reconstruction of quantum theory, ``An elementary system has the information carrying capacity of at most one bit'' \cite{dakic}. The difference between the classical prediction and the quantum reality uniquely distinguishes the quantum joint distribution from the classical joint distribution for the Bell spin states \cite{garg}.}  \label{SGExp}
\end{center}
\end{figure}

\begin{figure}
\begin{center}
\includegraphics [height = 50mm]{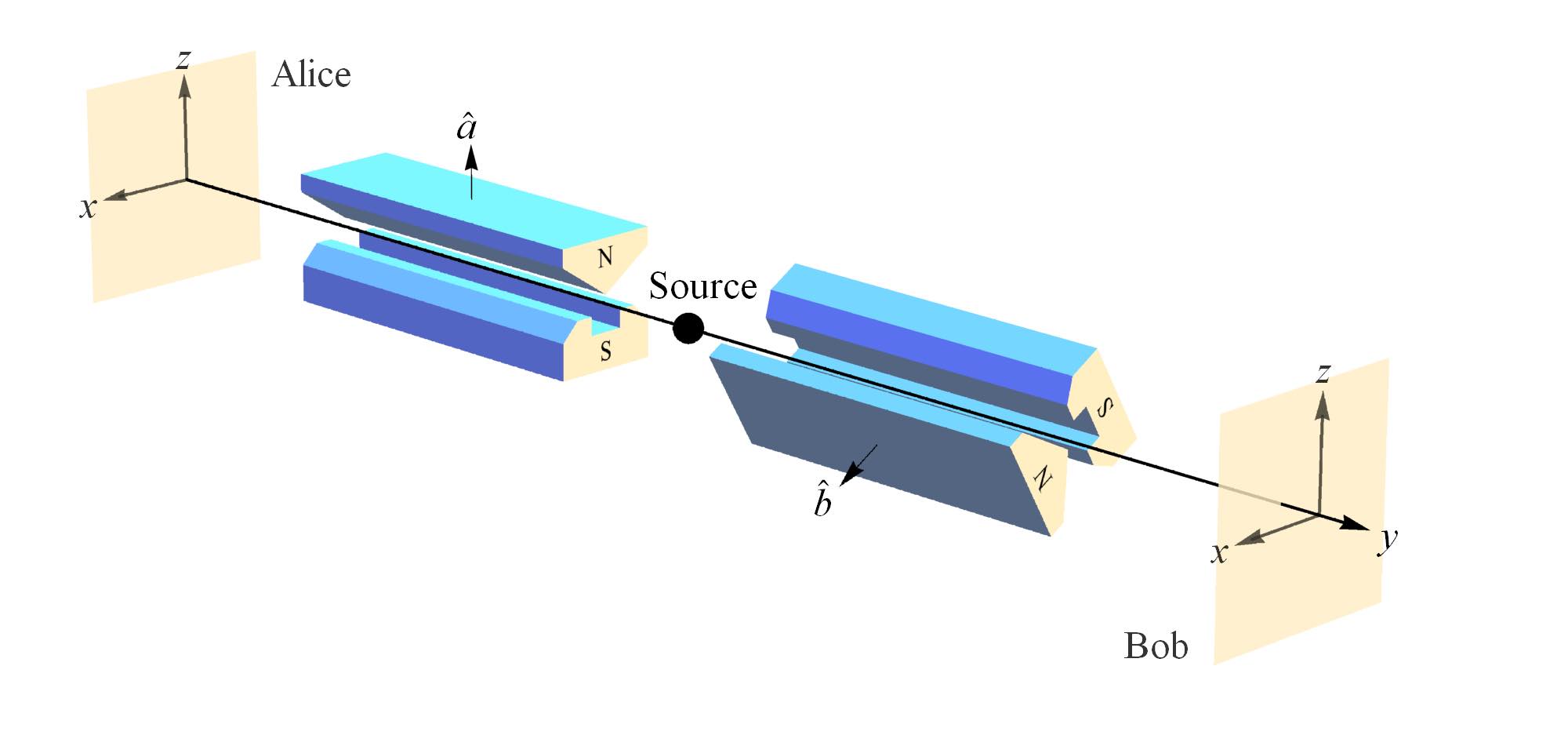}
\caption{Alice and Bob making spin measurements on a pair of spin-entangled particles with their Stern-Gerlach (SG) magnets and detectors in the $xz$-plane. Here Alice and Bob's SG magnets are not aligned so these measurements represent different reference frames. Since their outcomes satisfy Dakic and Brukner's Axiom 1 in all reference frames and satisfy explicit conservation of spin angular momentum in the same reference frame, they can only satisfy conservation of spin angular momentum on \textit{average} in different reference frames.} \label{EPRBmeasure}
\end{center}
\end{figure}

As Steven Weinberg points out, measuring an electron's spin via Stern-Gerlach (SG) magnets constitutes the measurement of ``a universal constant of nature, Planck's constant'' \cite[p. 3]{weinberg2017} (Figure \ref{SGExp}). So if NPRF applies equally here, everyone must measure the same value for Planck's constant \textit{h} regardless of their SG magnet orientations relative to the source, which like the light postulate is an empirical fact. By ``relative to the source'' of a pair of spin-entangled particles, we mean relative ``to the vertical in the plane perpendicular to the line of flight of the particles'' \cite[p. 943]{mermin1981} (Figure \ref{EPRBmeasure}). Here the possible spin outcomes $\pm\frac{\hbar}{2}$ represent a fundamental (indivisible) unit of information per Dakic and Brukner's first axiom in their reconstruction of quantum theory, ``An elementary system has the information carrying capacity of at most one bit'' \cite{dakic}. Thus, different SG magnet orientations relative to the source constitute different ``reference frames'' in QM just as different velocities relative to the source constitute different ``reference frames'' in SR. Borrowing from Einstein, NPRF might be stated \cite{einstein}:
\begin{quote}
    No one's ``sense experiences,'' to include measurement outcomes, can provide a privileged perspective on the ``real external world.'' 
\end{quote}
This is consistent with the notion of symmetries per Hicks \cite{hicks2019}:
\begin{quote}
    There are not two worlds in one of which I am here and in the other I am three feet to the left, with everything else similarly shifted. Instead, there is just this world and two mathematical descriptions of it. The fact that those descriptions put the origin at different places does not indicate any difference between the worlds, as the origin in our mathematical description did not correspond to anything in the world anyway. The symmetries tell us what structure the world does not have.
\end{quote}
That is, there is just one ``real external world'' harboring many, but always equal perspectives as far as the physics is concerned \cite{NMEntropy2020}.

\begin{figure}
\begin{center}
\includegraphics [height = 40mm]{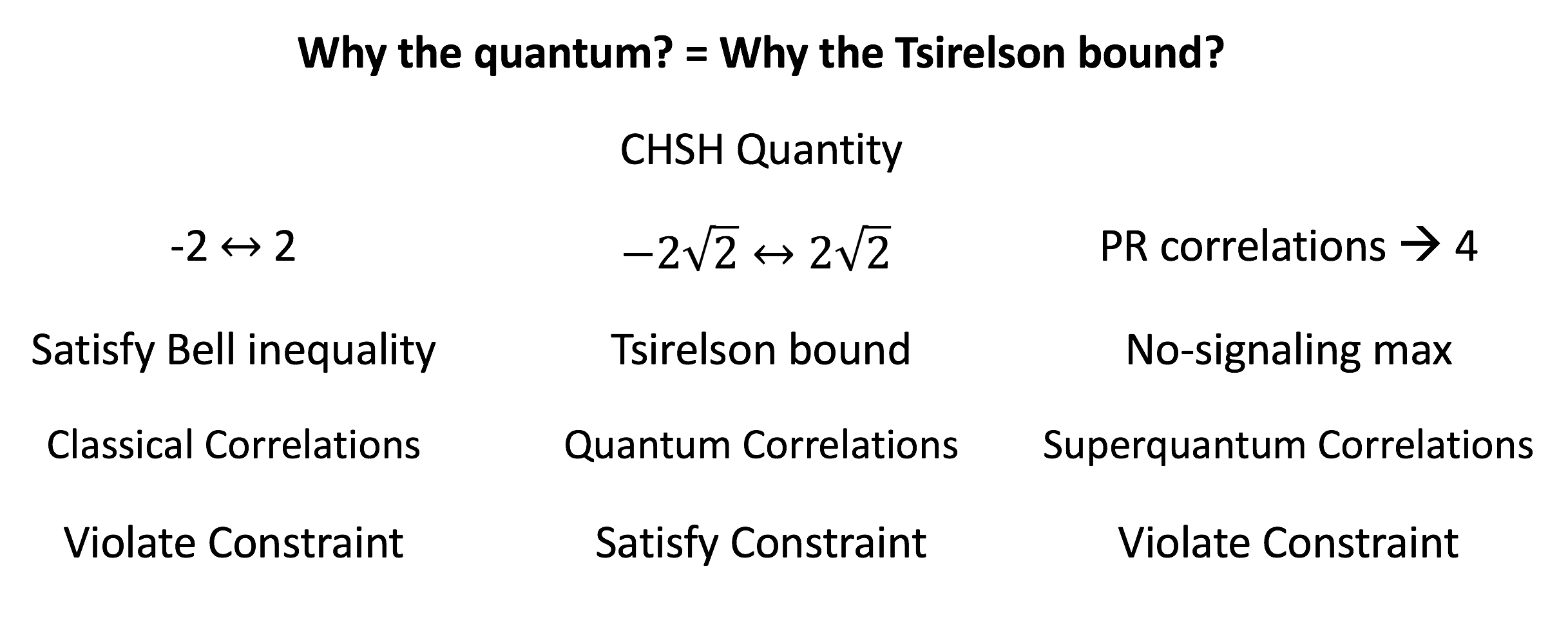}  \caption{\textbf{Answer to Bub's question, ``Why the Tsirelson bound?''} The ``constraint'' is conservation per no preferred reference frame.} \label{Summary}
\end{center}
\end{figure}

We have shown elsewhere that the quantum correlations and quantum states corresponding to the Bell states, which uniquely produce the Tsirelson bound for the Clauser–Horne–Shimony–Holt (CHSH) quantity, can be derived from conservation per NPRF \cite{TsirelsonBound2019}. Thus, Bell state entanglement is ultimately grounded in NPRF just as SR \cite{MerminChallenge}. As summarized in Figure \ref{Summary}, the quantum correlations responsible for the Tsirelson bound satisfy conservation per NPRF while both classical and superquantum correlations can violate this constraint. Therefore a principle explanation of Bell state entanglement and the Tsirelson bound that be stated in ``one clear, simple sentence'' \citep[p. 302]{fuchs1} is ``conservation per no preferred reference frame'' (Figure \ref{Summary}). 

What qualifies as a principle explanation versus constructive turns out to be a fraught and nuanced question \cite{felline} and we do not want to be sidetracked on that issue as such. Let us therefore state explicitly that what makes our explanation a principle one is that it is grounded directly in phenomenology, it is an adynamical and acausal explanation that involves adynamical global constraints as opposed to dynamical laws or causal mechanisms, and it is unifying with respect to QM and SR.

Let us also note that while contrary to certain others \cite{brownbook,brownpooley2006,norton2008,menon2019}, we are arguing that conservation per NPRF need not ever be discharged by a constructive explanation or interpretation. This is at least partially distinct from the question in SR for example, of whether facts about physical geometry are grounded in facts about dynamical fields or vice-versa. Furthermore, this principle explanation is consistent with any number of ``constructive interpretations'' of QM. For example, this principle explanation avoids the complaints about Bub's proposed principle explanation of QM leveled by Felline \cite{felline2018}. That is, the principle being posited herein does not require a solution to the measurement problem nor again does it necessarily beg for a constructive counterpart.

The Bell states are
\begin{equation}
\begin{aligned}
&|\psi_-\rangle = \frac{|ud\rangle \,- |du\rangle}{\sqrt{2}}\\
&|\psi_+\rangle = \frac{|ud\rangle + |du\rangle}{\sqrt{2}}\\
&|\phi_-\rangle = \frac{|uu\rangle \,- |dd\rangle}{\sqrt{2}}\\
&|\phi_+\rangle = \frac{|uu\rangle + |dd\rangle}{\sqrt{2}}\\ \label{BellStates}
\end{aligned}
\end{equation}
in the eigenbasis of $\sigma_z$. The first state $|\psi_-\rangle$ is called the ``spin singlet state'' and it represents a total conserved spin angular momentum of zero ($S = 0$) for the two particles involved. The other three states are called the ``spin triplet states'' and they each represent a total conserved spin angular momentum of one ($S = 1$, in units of $\hbar = 1$ for spin-$\frac{1}{2}$ particles). In all four cases, the entanglement represents the conservation of spin angular momentum for the process creating the state. 

If Alice is making her spin measurement $\sigma_1$ in the $\hat{a}$ direction and Bob is making his spin measurement $\sigma_2$ in the $\hat{b}$ direction, we have
\begin{equation}
\begin{aligned}
    &\sigma_1 = \hat{a}\cdot\vec{\sigma}=a_x\sigma_x + a_y\sigma_y + a_z\sigma_z \\
    &\sigma_2 = \hat{b}\cdot\vec{\sigma}=b_x\sigma_x + b_y\sigma_y + b_z\sigma_z \\ \label{sigmas}
\end{aligned}
\end{equation}
The correlation functions are given by \cite{MerminChallenge}
\begin{equation}
\begin{aligned}
&\langle\psi_-|\sigma_1\sigma_2|\psi_-\rangle = &-a_xb_x - a_yb_y - a_zb_z\\
&\langle\psi_+|\sigma_1\sigma_2|\psi_+\rangle = &a_xb_x + a_yb_y - a_zb_z\\
&\langle\phi_-|\sigma_1\sigma_2|\phi_-\rangle = &-a_xb_x + a_yb_y + a_zb_z\\
&\langle\phi_+|\sigma_1\sigma_2|\phi_+\rangle = &a_xb_x - a_yb_y + a_zb_z\\ \label{gencorrelations}
\end{aligned}
\end{equation}

The spin singlet state is invariant under all three SU(2) transformations meaning we obtain opposite outcomes ($\frac{1}{2}$ $ud$ and $\frac{1}{2}$ $du$) for SG magnets at any $\hat{a}=\hat{b}$ (Figures \ref{SGExp} \& \ref{EPRBmeasure}) and a correlation function of $-\cos(\theta)$ in any plane of physical space, where $\theta$ is the angle between $\hat{a}$ and $\hat{b}$ (Eq. (\ref{gencorrelations})). We see that the conserved spin angular momentum ($S = 0$), being directionless, is conserved in any plane of physical space. Again, $\hat{a}=\hat{b}$ means Alice and Bob are in the same reference frame.

The invariance of each of the spin triplet states under its respective SU(2) transformation in Hilbert space represents the SO(3) invariant conservation of spin angular momentum $S = 1$ for each of the planes $xz$ ($|\phi_+\rangle$), $yz$ ($|\phi_-\rangle$), and $xy$ ($|\psi_+\rangle$) in physical space. Specifically, when the SG magnets are aligned (the measurements are being made in the same reference frame) anywhere in the respective plane of symmetry the outcomes are always the same ($\frac{1}{2}$ $uu$ and $\frac{1}{2}$ $dd$). It is a planar conservation and our experiment would determine which plane. If you want to model a conserved $S = 1$ for some other plane, you simply create a superposition, i.e., expand in the spin triplet basis. And in that plane, you're right back to the mystery of Bell state entanglement per conserved spin angular momentum via a correlation function of $\cos(\theta)$, as with any of the spin triplet states (Eq. (\ref{gencorrelations})). 

We will explain the spin singlet state correlation function, since the spin triplet state correlation function is analogous. That we have opposite outcomes when Alice and Bob are in the same reference frame is not difficult to understand via conservation of spin angular momentum, because Alice and Bob's measured values of spin angular momentum cancel directly when $\hat{a}=\hat{b}$ (Figure \ref{EPRBmeasure}). But, when Bob's SG magnets are rotated by $\theta$ relative to Alice's SG magnets, we need to clarify the situation.

We have two subsets of data, Alice's set (with SG magnets at angle $\alpha$) and Bob's set (with SG magnets at angle $\beta$). They were collected in $N$ pairs (data events) with Bob's(Alice's) SG magnets at $\alpha - \beta = \theta$ relative to Alice's(Bob's). We want to compute the correlation function for these $N$ data events which is

\begin{equation}\langle \alpha,\beta \rangle =\frac{(+1)_A(-1)_B + (+1)_A(+1)_B + (-1)_A(-1)_B + ...}{N}\end{equation}
Now partition the numerator into two equal subsets per Alice's equivalence relation, i.e., Alice's $+1$ results and Alice's $-1$ results

\begin{equation}\langle \alpha,\beta \rangle =\frac{(+1)_A(\sum \mbox{BA+})+(-1)_A(\sum \mbox{BA-})}{N} \label{correl2}\end{equation}
where $\sum \mbox{BA+}$ is the sum of all of Bob's results (event labels) corresponding to Alice's $+1$ result (event label) and $\sum \mbox{BA-}$ is the sum of all of Bob's results (event labels) corresponding to Alice's $-1$ result (event label). Notice this is all independent of the formalism of QM. Next, rewrite Eq. (\ref{correl2}) as
\begin{equation}\langle \alpha,\beta \rangle = \frac{1}{2}(+1)_A\overline{BA+} + \frac{1}{2}(-1)_A\overline{BA-} \label{consCorrel}\end{equation}
with the overline denoting average. Notice that to understand the quantum correlation responsible for Bell state entanglement, we need to understand the origins of $\overline{BA+}$ and $\overline{BA-}$ for the Bell states. We now show what that is for the spin singlet state \cite{unnik2005}, the spin triplet states are analogous in their respective symmetry planes \cite{MerminChallenge}.

\begin{figure}
\begin{center}
\includegraphics [height = 25mm]{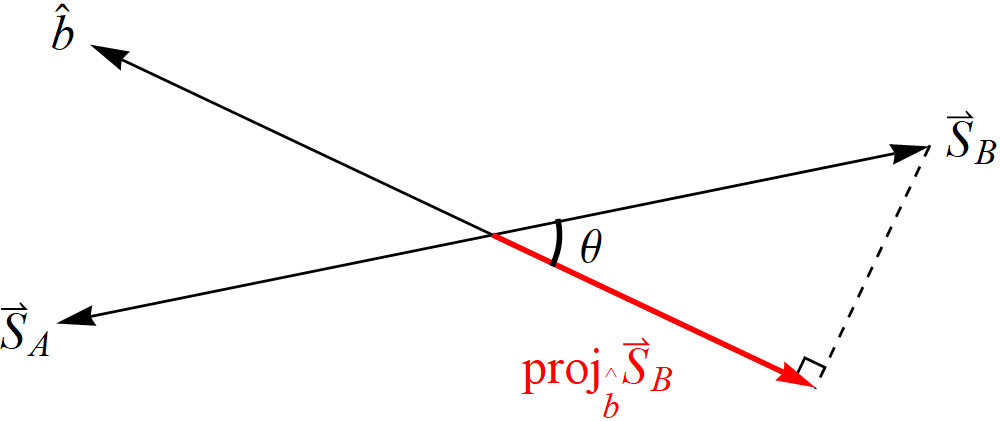}
\caption{The spin angular momentum of Bob's particle $\vec{S}_B = -\vec{S}_A$ projected along his measurement direction $\hat{b}$. This does \textit{not} happen with spin angular momentum.} \label{Projection1}
\end{center}
\end{figure}

\begin{figure}
\begin{center}
\includegraphics [height = 25mm]{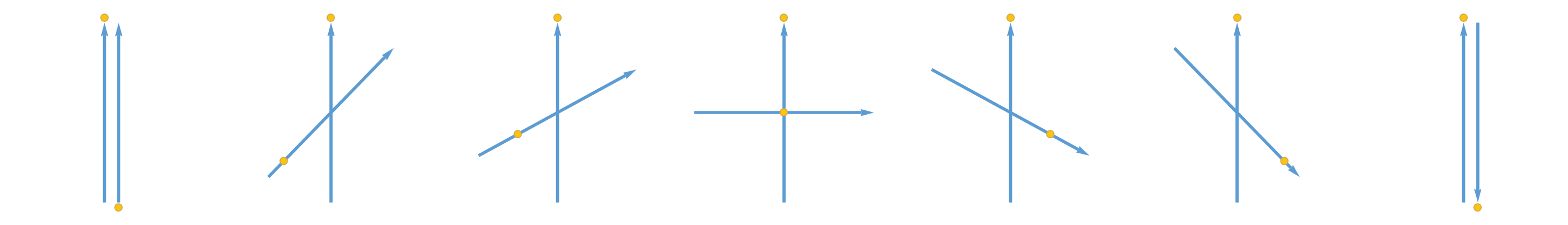}  
\caption{\textbf{Average View for the Spin Singlet State}. Reading from left to right, as Bob rotates his SG magnets relative to Alice's SG magnets for her $+1$ outcome, the average value of his outcome varies from $-1$ (totally down, arrow bottom) to 0 to +1 (totally up, arrow tip). This obtains per conservation of spin angular momentum on average in accord with no preferred reference frame. Bob can say exactly the same about Alice's outcomes as she rotates her SG magnets relative to his SG magnets for his $+1$ outcome. That is, their outcomes can only satisfy conservation of spin angular momentum \textit{on average} in different reference frames, because they only measure $\pm 1$, never a fractional result. Thus, just as with the light postulate of SR, we see that no preferred reference frame leads to a counterintuitive result. Here it requires quantum outcomes $\pm 1 \left(\frac{\hbar}{2}\right)$ for all measurements and that leads to the mystery of ``average-only'' conservation.} \label{AvgViewSinglet}
\end{center}
\end{figure}

In classical physics, one would say the projection of the spin angular momentum vector of Alice's particle $\vec{S}_A = +1\hat{a}$ along $\hat{b}$ is $\vec{S}_A\cdot\hat{b} = +\cos({\theta})$ where again $\theta$ is the angle between the unit vectors $\hat{a}$ and $\hat{b}$. That's because the prediction from classical physics is that all values between $+1 \left(\frac{\hbar}{2}\right)$ and $-1 \left(\frac{\hbar}{2}\right)$ are possible outcomes for a spin measurement (Figure \ref{SGExp}). From Alice's perspective, had Bob measured at the same angle, i.e., $\beta = \alpha$, he would have found the spin angular momentum vector of his particle was $\vec{S}_B = -\vec{S}_A = -1\hat{a}$, so that $\vec{S}_A + \vec{S}_B = \vec{S}_{Total} = 0$. Since he did not measure the spin angular momentum of his particle at the same angle, he should have obtained a fraction of the length of $\vec{S}_B$, i.e., $\vec{S}_B\cdot\hat{b} = -1\hat{a}\cdot\hat{b} = -\cos({\theta})$ (Figure \ref{Projection1}; this also follows from counterfactual spin measurements on the single-particle state \cite{boughn}). Of course, Bob only ever obtains $+1$ or $-1$ per NPRF, but suppose that Bob's outcomes \textit{average} $-\cos({\theta})$ (Figure \ref{AvgViewSinglet}). This means 
\begin{equation}\overline{BA+} = -\cos(\theta) \label{AvgPlus1}\end{equation}
Likewise, for Alice's $(-1)_A$ results we have
\begin{equation}\overline{BA-} = \cos(\theta) \label{AvgMinus1}\end{equation}
Putting these into Eq. (\ref{consCorrel}) we obtain
\begin{equation}\langle \alpha,\beta \rangle = \frac{1}{2}(+1)_A(-\cos(\theta)) + \frac{1}{2}(-1)_A(\cos(\theta)) = -\cos(\theta) \label{correlEnd}\end{equation}
which is precisely the correlation function given by QM for the spin singlet state. Notice that Eqs. (\ref{AvgPlus1}) \& (\ref{AvgMinus1}) are mathematical facts for obtaining the quantum correlation function, we are simply motivating these facts via conservation of spin angular momentum in accord with the SU(2) Bell state invariances. Of course, Bob could partition the data according to his equivalence relation (per his reference frame) and claim that it is Alice who must average her results (obtained in her reference frame) to conserve angular momentum (Figure \ref{AvgViewSinglet}).

\begin{figure}
\begin{center}
\includegraphics [width=\textwidth]{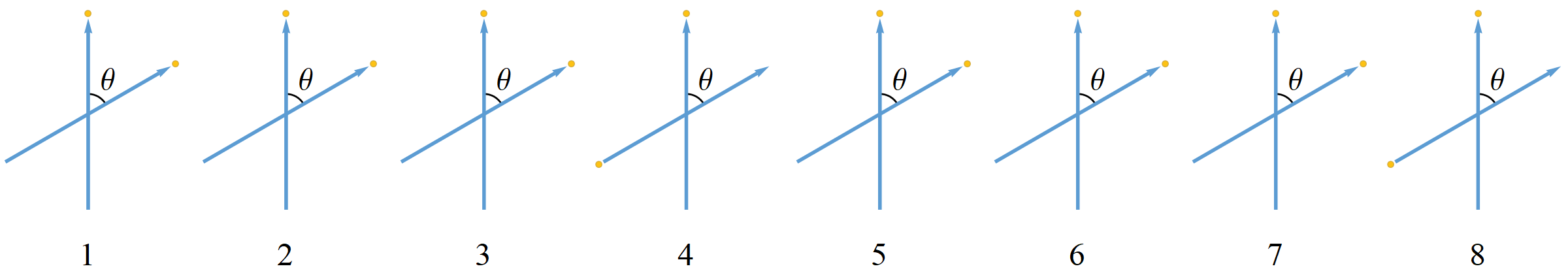} 
\caption{A spatiotemporal ensemble of 8 experimental trials for the spin triplet states showing Bob’s outcomes corresponding to Alice's $+1$ outcomes when $\theta = 60^\circ$. Spin angular momentum is not conserved in any given trial, because there are two different measurements being made, i.e., outcomes are in two different reference frames, but it is conserved on average for all 8 trials (six up outcomes and two down outcomes average to $\cos{60^\circ}=\frac{1}{2}$). It is impossible for spin angular momentum to be conserved explicitly in any given trial since the measurement outcomes are binary (quantum) with values of $+1$ (up) or $-1$ (down) per no preferred reference frame and explicit conservation of spin angular momentum in different reference frames would require a fractional outcome for Alice and/or Bob.} \label{4Dpattern}
\end{center}
\end{figure}

We posit that the reason we have average-only conservation in different reference frames is ultimately due to NPRF. To motivate NPRF for the Bell states, consider the empirical facts. First, Bob and Alice both measure $\pm 1 \left(\frac{\hbar}{2}\right)$ for all SG magnet orientations, i.e. in all reference frames. In order to satisfy conservation of spin angular momentum for any given trial when Alice and Bob are making different measurements, i.e., when they are in different reference frames, it would be necessary for Bob or Alice to measure some fraction, $\pm \cos(\theta)$. For example, if Alice measured $+1$ at $\alpha = 0$ for an $S = 1$ state (in the plane of symmetry) and Bob made his measurement (in the plane of symmetry) at $\beta = 60^\circ$, then Bob's outcome would need to be $\frac{1}{2}$ (Figure \ref{4Dpattern}). In that case, we would know that Alice measured the ``true'' spin angular momentum of her particle (and therefore the ``true'' value of Planck's constant) while Bob only measured a component of the ``true'' spin angular momentum for his particle. Thus, Alice's SG magnet orientation would definitely constitute a ``preferred reference frame.'' 

But, this is precisely what does \textit{not} happen. Alice and Bob both always measure $\pm 1 \left(\frac{\hbar}{2}\right)$, no fractions, in accord with NPRF for the measurement of Planck's constant. And, this fact alone distinguishes the quantum joint distribution from the classical joint distribution \cite{garg} (Figure \ref{SGExp}). Therefore, the average-only conservation responsible for the correlation function for the Bell states is actually conservation resulting from NPRF.

If Alice is moving at velocity $\vec{V}_a$ relative to a light source, then she measures the speed of light from that source to be \textit{c} ($=\frac{1}{\sqrt{\mu_o\epsilon_o}}$, as predicted by Maxwell's equations). If Bob is moving at velocity $\vec{V}_b$ relative to that same light source, then he measures the speed of light from that source to be \textit{c}. Here ``reference frame'' refers to the relative motion of the observer and source, so all observers who share the same relative velocity with respect to the source occupy the same reference frame. NPRF in this context means all measurements produce the same outcome \textit{c}. 

As a consequence of this constraint we have time dilation and length contraction, which are then reconciled per NPRF via the relativity of simultaneity. That is, Alice and Bob each partition spacetime per their own equivalence relations (per their own reference frames), so that equivalence classes are their own surfaces of simultaneity. If Alice's equivalence relation over the spacetime events yields the ``true'' partition of spacetime, then Bob must correct his lengths and times per length contraction and time dilation. Of course, the relativity of simultaneity says that Bob's equivalence relation is as valid as Alice's per NPRF. 

This is completely analogous to QM, where Alice and Bob each partition the data per their own equivalence relations (per their own reference frames), so that equivalence classes are their own $+1$ and $-1$ data events. If Alice's equivalence relation over the data events yields the ``true'' partition of the data, then Bob must correct (average) his results per average-only conservation. Of course, NPRF says that Bob's equivalence relation is as valid as Alice's, which we might call the ``relativity of data partition'' (Table \ref{tab:SRvsQM}).

\begin{table}[htbp]
    \center
    \begin{tabularx}{1.0\textwidth} { 
  | >{\raggedright\arraybackslash}X 
  | >{\raggedright\arraybackslash}X | }
    \hline
        {\bf Special Relativity} & {\bf Quantum Mechanics} \\
        \hline
        Empirical Fact: Alice and Bob both measure $c$, regardless of their motion relative to the source  & Empirical Fact: Alice and Bob both measure $\pm 1 \left(\frac{\hbar}{2} \right)$, regardless of their SG orientation relative to the source \\
        \hline
        Alice(Bob) says of Bob(Alice): Must correct time and length measurements  &  Alice(Bob) says of Bob(Alice): Must average results \\ 
        \hline
        NPRF: Relativity of simultaneity & NPRF: Relativity of data partition \\
        \hline
    \end{tabularx}
    \caption{\textbf{Comparing SR with QM according to no preferred reference frame (NPRF)}.}
    \label{tab:SRvsQM}
\end{table}

Thus, the mysteries of SR (time dilation and length contraction) ultimately follow from the same principle as Bell state entanglement, i.e., no preferred reference frame. So, if one accepts SR's principle explanation of time dilation and length contraction, then they should have no problem accepting conservation per NPRF as a principle explanation of Bell state entanglement. Thus, the relativity principle (NPRF) is a unifying principle for non-relativistic QM and SR, thereby addressing the desideratum of QIT in general and answering Bub's question specifically.

Despite the fact that this principle explanation supplies a unifying framework for both non-relativistic QM and SR, some might demand a constructive explanation with its corresponding ``knowledge of how things in the world work, that is, of the mechanisms (often hidden) that produce the phenomena we want to understand'' \cite[p. 15]{salmon1993}. This is ``the causal/mechanical view of scientific explanation'' per Salmon \cite[p. 15]{salmon1993}. Thus, as with SR, not everyone will consider our principle account to be explanatory since, ``By its very nature such a theory-of-principle explanation will have nothing to say about the reality behind the phenomenon'' \cite[p. 331]{balashov}. As stated by Brown \cite[p. 76]{brownpooley2006}:
\begin{quote}
    What has been shown is that rods and clocks must behave in quite particular ways in order for the two postulates to be true together. But this hardly amounts to an explanation of such behaviour. Rather things go the other way around. It is because rods and clocks behave as they do, in a way that is consistent with the relativity principle, that light is measured to have the same speed in each inertial frame.
\end{quote}

In other words, the assumption is that the true or fundamental ``explanation'' of Bell state entanglement must be a constructive one in the sense of adverting to causal mechanisms like fundamental physical entities such as particles or fields and their dynamical equations of motion. Notice that while our account of SR is in terms of fundamental principle explanation, that does not necessarily make it a ``geometric'' interpretation of SR. For example, nothing we've said commits us to the claim that if one were to remove all the matter-energy out of the universe there would be some geometric structure remaining such as Minkowski spacetime. Furthermore, there is nothing inherently geometric about our principle explanation of Bell state entanglement in particular or of NPRF in general.

Of course we do not have a no-go argument that our principle explanation will never be subsumed by a constructive one. However, especially in light of the unifying nature of our principle explanation, we think it is worth considering the possibility that principle explanation is fundamental in these cases and perhaps others \cite{ourbook,TsirelsonBound2019,MerminChallenge}. We think this is especially reasonable in light of the current impasse in both QIT-based explanations of QM phenomena and in attempts at constructive interpretations. Essentially, we are in a situation with QM that Einstein found himself in with SR \cite[pp. 51-52]{einstein1949}:
\begin{quote}
By and by I despaired of the possibility of discovering the true laws by means of constructive efforts based on known facts. The longer and the more despairingly I tried, the more I came to the conviction that only the discovery of a universal formal principle could lead us to assured results. The example I saw before me was thermodynamics.
\end{quote}

Thus we are offering a competing account of quantum entanglement for any interpretation that fundamentally explains entanglement in the constructive sense. As Einstein said, this gives us the advantage of ``logical perfection and security of the foundations'' as our principle account could be true across a number of different constructive interpretations. And, the principle we offer, NPRF, is a unifying principle for non-relativistic QM and SR that holds throughout physics \cite{NMEntropy2020}. As Pauli once stated \cite[p. 33]{Pauli1971}:
\begin{quote}
`Understanding' probably means nothing more than having whatever ideas and concepts are needed to recognize that a great many different phenomena are part of a coherent whole.
\end{quote}

Per Hicks \cite{hicks2019}, NPRF is a principle that is accessible (``because it is simple'') and whence we can ``infer lots of truths.'' Inferring ``lots of truths'' implies a unifying principle is superior to its subsumed constituents, since it implies (at minimum) more truths than any proper subset of its subsumed constituents. The point is, we are hypothesizing that the SO(3) symmetry with average-only conservation as an explanation of Bell state entanglement, and Lorentz symmetry with relativity of simultaneity as an explanation of length contraction and time dilation, are expressions of a deeper truth, NPRF, with seemingly disparate multiple physical consequences. It has been suggested that perhaps other unresolved phenomena in physics might be explained in a similar fashion \cite{ourbook}. 

The bottom line is that a compelling constraint (who would argue with conservation per NPRF?) explains Bell state entanglement without any obvious corresponding `dynamical/causal influence' or hidden variables to account for the results on a trial-by-trial basis. By accepting this principle explanation as fundamental, the lack of a compelling, consensus constructive explanation is not a problem. This is just one of many mysteries in physics created by dynamical and causal biases that can be resolved by constraint-based thinking \cite{ourbook}.


\section{\label{wigner}Principle Explanation of Wigner's Friend} 
We introduce the quantum mechanical gedanken experiment called ``Wigner's friend'' using Healey's version \cite{healey} of Frauchiger \& Renner's version \cite{FR} of Wigner's original version \cite{wigner}. The whole point of the Wigner's friend scenario is that someone (Wigner in the original story) makes a quantum measurement of someone else (Wigner's friend) who made a measurement of some quantum system. For that to be possible, Wigner's friend must be isolated (screened off) from Wigner and the rest of the universe. Being screened off means Wigner's friend cannot share any classical information with the universe. This introduces crucial (but often ignored) technical and conceptual difficulties. 

Technically, one would have to keep Wigner's friend and his entire lab from interacting with the universe, e.g., no exchange of photons. That's certainly beyond anything we can do now, but more importantly, this means the classical information possessed by the friend\footnote{The classical information could just as well be contained in a computer with no human agent involved.} with lab while screened off is not accessible to Wigner or anyone else in the universe. Since the universe is precisely all shared, self-consistent classical information, the friend's classical information while screened off, being unshared/unaccessible to everyone else, is not even part of the universe and therefore not a part of objective reality. And that means, among other things, that there is no way to establish relative coordinate directions between Wigner's frame and his friend's frame. The alignment of Cartesian frames and synchronization of clocks between observers is already a problem that must be overcome to transfer information via quantum systems \cite{Bartlett2007}. That is because we must first be able to relate Hilbert space vectors involving, say, polarizer orientations between the two isolated frames in order to use the formalism of QM. Thus, after being measured by Wigner, it is impossible for the friend's classical results to contradict the shared classical information of the universe, as required to refute objective reality.

Conceptually, if Wigner's friend is measuring $\hat{x}$ with the eigenbasis $|heads\rangle$ and $|tails\rangle$ (a ``quantum coin flip''), it must be possible for Wigner to measure $\hat{w}$ with the eigenbasis $|heads\rangle - |tails\rangle$ and $|heads\rangle + |tails\rangle$, even though we cannot imagine what that means in terms of a coin flip\footnote{This is another crucial point that is ignored in analyses of Wigner's friend, but as we will see in section \ref{WFexperiment}, Proietti et al. \cite{proietti} have shed some light on it.}. In QM, every Hilbert space basis rotated from the eigenbasis of some measurement operator is the eigenbasis of some other measurement operator and therefore constitutes something we can measure, e.g., Stern-Gerlach magnets or polarizers rotated in space giving rise to transformed eigenbases in Hilbert space. 

As we will show, the tacit introduction of these inconsistencies is what leads to the inconsistencies (inconsistent, shared classical information in the form of shared measurement outcomes) associated with the Wigner's friend experiment. And, as Baumann \& Wolf (BW) show \cite{baumann} and we will introduce, there are other approaches to QM which do not suffer such inconsistencies. 

Thus, the bottom-line question for the Wigner's friend experiment is whether or not a person (such as Wigner's friend) making a quantum measurement can themselves be treated consistently as a quantum system by someone else (such as Wigner). As we will see, the answer is ``yes,'' as long as the friend does not share classical information with the universe while screened off, i.e., while being treated as a quantum system. So, there is no size limit to the applicability of QM as some have asserted based on FR. The problem is, the way Wigner's friend is typically cast, the friend employs a measurement-update rule\footnote{This is sometimes associated with ``wavefunction collapse.''} while Wigner assumes the friend and his lab (to include their measurement records and memories) continue to evolve unitarily per the Schr{\"o}dinger equation. BW call this ``subjective collapse'' and show that even in the simplest version of Wigner's friend, the sharing of inconsistent classical information can result. We will provide a similar example below. According to our constraint-based explanation, since this possibility allows for self-inconsistent shared classical information, the answer to this bottom-line question is ``no'' under these circumstances. The whole point of the constraint in the universe (such as Einstein's equations) is to ensure that the shared classical information composing the universe is self-consistent. We begin with a presentation of Wigner's friend per Healey showing how subjective collapse leads to the sharing of inconsistent classical information. Then we will show how proper treatments per the ``standard'' and ``relative-state'' formalisms for QM contain no such inconsistency.

There are four agents in this story -- Xena who makes a quantum measurement $\hat{x}$ on quantum state c in her lab X and then sends a quantum state s to Yvonne, Yvonne who makes a quantum measurement $\hat{y}$ on s in her lab Y, Zeus who makes a quantum measurement $\hat{z}$ or $\hat{x}$ on X pertaining to Xena's $\hat{x}$ measurement, and Wigner who makes a quantum measurement $\hat{w}$ or $\hat{y}$ on Y pertaining to Yvonne's $\hat{y}$ measurement. So, we have two ``Wigners,'' i.e., Zeus and Wigner, and two ``Wigner's friends,'' i.e., Xena (Zeus's friend) and Yvonne (Wigner's friend). The first assumption of FR is that it is possible for Xena and Yvonne to behave as quantum systems for Zeus and Wigner to measure, i.e., there are no size restrictions on what can behave quantum mechanically. Again, we agree that it is possible to screen off Xena and Yvonne with the caveat that, in Bub's language, those systems being treated quantum mechanically per non-Boolean algebra are not exchanging classical information with the universe per Boolean algebra. We now explore the implications of violating that caveat via subjective collapse as in FR. The starting state c for Xena is

\begin{equation} 
\frac{1}{\sqrt{3}}| heads \rangle + \frac{\sqrt{2}}{\sqrt{3}}| tails \rangle \label{c}
\end{equation}
The eigenbasis for Xena's $\hat{x}$ measurement is simply $| heads \rangle$ and $| tails \rangle$ with eigenvalues heads and tails, respectively. If the outcome of her measurement is heads, she sends state s

\begin{equation} 
| - \rangle \label{s+}
\end{equation}
to Yvonne. If the outcome of her measurement is tails, she sends state s

\begin{equation} 
\frac{1}{\sqrt{2}}\left(| + \rangle + | - \rangle \right) \label{s-}
\end{equation}
to Yvonne. Again, if $| + \rangle$ and $| - \rangle$ refer to orientations in space, e.g., polarizers or Stern-Gerlach magnets, their meaning between Xena and Yvonne is problematic, since their labs are isolated from one another. The sharing of such classical information is crucial for the modeling of the spacetime universe as we have defined it per the very meaning of the Hilbert space structure, so ignoring this point is to introduce an inconsistency. But, as with the original papers, we will proceed to find its implications. 

The second assumption of FR is that there is only one outcome for a quantum measurement. So, for example, Xena does not measure both heads and tails and send both versions of state s. Now, assuming the subjective-collapse model, Xena and Yvonne's labs are behaving quantum mechanically (evolving unitarily) according to Zeus and Wigner, so they are entangled in the state

\begin{equation} 
|\Psi \rangle = \frac{1}{\sqrt{3}}\left(| heads \rangle | - \rangle + | tails \rangle | + \rangle + |tails \rangle | - \rangle \right)\label{Eq13}
\end{equation}
per Eqs. (\ref{c}), (\ref{s+}) and (\ref{s-}) for Zeus and Wigner (this is Eq. (13) in Healey's paper). Assuming Xena and Yvonne's labs evolve unitarily means Zeus and Wigner can make measurements of Xena and Yvonne's labs in any rotated Hilbert space basis they choose. In addition to the measurement $\hat{x}$ with eigenbasis $|heads\rangle$ and $|tails\rangle$, Zeus has the option of measuring $\hat{z}$ with eigenbasis

\begin{equation}
\begin{split}
|OK \rangle_Z = &\frac{1}{\sqrt{2}} \left(|heads \rangle - |tails \rangle \right)\\
|fail \rangle_Z = &\frac{1}{\sqrt{2}} \left(|heads \rangle + |tails \rangle \right)\\\label{zeus}
\end{split}
\end{equation}
And, in addition to the measurement $\hat{y}$ with eigenbasis $|+\rangle$ and $|-\rangle$, Wigner has the option of measuring $\hat{w}$ with eigenbasis

\begin{equation}
\begin{split}
|OK \rangle_W = &\frac{1}{\sqrt{2}} \left(|+ \rangle - |- \rangle \right)\\
|fail \rangle_W = &\frac{1}{\sqrt{2}} \left(|+ \rangle + |- \rangle \right)\\\label{Wigner}
\end{split}
\end{equation}
Again, this introduces another inconsistency with the universe of shared, self-consistent classical information, i.e., the classical information must be intelligible. First, we have already pointed out how the relative orientations of Stern-Gerlach magnets or polarizers are needed to make sense of Wigner's measurements of Yvonne's lab and that this constitutes shared classical information forbidden by the assumption that Yvonne's lab is screened off from Wigner. Second, we also have a problem of intelligibility between Zeus and Xena as mentioned above. That is, the eigenbasis $|heads\rangle - |tails\rangle$ and $|heads\rangle + |tails\rangle$ makes no sense and without that understanding, it is impossible to create a spacetime universe experimental configuration of classical information corresponding to this Hilbert space basis\footnote{See section \ref{WFexperiment} concerning progress on this question from Proietti et al. \cite{proietti}.}. Now let us continue and show how these inconsistencies play out.

In any given trial of the experiment, since Xena and Yvonne have definite outcomes duly recorded and memorized, there is classical information as to which of the three possible outcomes in Eq. (\ref{Eq13}) was actually instantiated, i.e., Xena obtained heads and Yvonne obtained --1, Xena obtained tails and Yvonne obtained +1, or Xena obtained tails and Yvonne obtained --1. But, it is easy to show that this classical information is not consistent with the entangled state $|\Psi\rangle$. Essentially, as Bub points out \cite{bubforthcoming}, this is just to say that mere ignorance about classical information does not constitute a quantum system. Again, in Bub's language, classical information is Boolean while quantum information is non-Boolean.

Suppose Zeus measures $\hat{z}$ and obtains OK (eigenvalue for $|OK\rangle_Z$), which can certainly happen since Xena measured either heads or tails. In other words, since we are assuming $|\Psi\rangle$ is the quantum state being measured by Zeus and Wigner for any of the definite configurations for Xena and Yvonne, and the projection of $|\Psi\rangle$ onto $|OK \rangle_Z$ is non-zero, it must be possible for Zeus to obtain OK for a $\hat{z}$ measurement for any prior definite configuration for Xena and Yvonne. That is, each of the three individual possibilities of heads and --1 or tails and +1 or tails and --1 is compatible with an OK outcome for Zeus's $\hat{z}$ measurement. But when Zeus obtains OK, Wigner must obtain +1 for a $\hat{y}$ measurement, since $\langle\Psi\mid OK\rangle_Z = -\frac{1}{\sqrt{6}}\langle +|$, which rules out the possibility that the prior configuration of Xena and Yvonne was heads and --1 or tails and --1, respectively, and that contradicts our assumption that $|\Psi\rangle$ is the quantum state for Zeus and Wigner for any of the definite configurations for Xena and Yvonne prior to Zeus and Wigner's measurements. Obviously, our mistake is to assume classical definiteness for a quantum system. You can see that we have a QM interference effect by rewriting Eq. (\ref{Eq13}) as

\begin{equation} 
|\Psi \rangle = \frac{1}{\sqrt{3}}\left(\sqrt{2}|fail \rangle_Z | - \rangle + |tails \rangle | +\rangle \right)\label{Bub1}
\end{equation}

Likewise, suppose Wigner first measures $\hat{w}$ and obtains OK (eigenvalue for $|OK\rangle_Y$), which can certainly happen since Yvonne measured either +1 or --1, i.e., each of the three individual possibilities of heads and --1 or tails and +1 or tails and --1 is compatible with an OK outcome for Wigner’s $\hat{w}$ measurement. But, when Wigner obtains OK, Zeus must obtain heads if he measures $\hat{x}$, since $\langle\Psi\mid OK\rangle_Y = -\frac{1}{\sqrt{6}}\langle heads|$, which rules out tails and +1 or tails and --1 as possible prior configurations for Xena and Yvonne, respectively, contrary to our assumption. Again, QM interference is at work, which you can see by rewriting Eq. (\ref{Eq13}) as

\begin{equation} 
|\Psi \rangle = \frac{1}{\sqrt{3}}\left(| heads \rangle | - \rangle + | tails \rangle |fail \rangle_W \right)\label{Bub2}
\end{equation}
Now let us show explicitly what kind of shared, self-inconsistent classical information can result for the inconsistent assumptions involved in this subjective-collapse experiment.  

Suppose Xena measures the state 

\begin{equation} 
\frac{1}{\sqrt{2}}\left(| heads \rangle + | tails \rangle \right) \label{Xena}
\end{equation}
in the basis $| heads \rangle, | tails \rangle$. Further, suppose that if Xena obtains heads, she sends the state $| heads \rangle$ to Wigner who likewise does a measurement in the basis $| heads \rangle, | tails \rangle$. Conversely, if Xena measures tails, she sends the state $| tails \rangle$ to Wigner who again does a measurement in the basis $| heads \rangle, | tails \rangle$. This constitutes sending classical information of course, which is essentially what is done in FR's approach. Now, Zeus passes Xena and her lab through a `heads-tails polarizer' $\left(| heads \rangle + | tails \rangle \right)$ and then does a measurement in the the basis $| heads \rangle, | tails \rangle$. Of course, it is entirely possible that Zeus's final measurement will yield tails. Keep in mind that Zeus's outcome is classical information about Xena's entire recorded history to include Xena's memories of the entire process. Therefore, Xena's records and memories will show that she measured tails and sent $| tails \rangle$ to Wigner. Consequently, Wigner must have a classically written record of tails for his outcome. But, of course, Wigner's classical information is heads, so we do not have a self-consistent collection of classical information being shared between Xena, Zeus, and Wigner. BW call this a ``scientific contradiction'' and state that it ``must not arise for a scientific theory.'' We concur of course. So, how would the standard formalism of QM deal with this?

In the standard formalism per BW, everyone agrees that all measurements produce a collapse, i.e., measurement update is objective. Therefore, Xena and Yvonne's measurement outcomes constitute classical information that cannot be treated quantum mechanically if it is going to be shared as part of the universe. [Of course, one could argue that ``unshared classical information'' is an oxymoron, but that is a semantic point with ontological implications that we will not engage here.] Recall, a screened-off system is not part of the universe/objective reality by definition. However, that does not mean we cannot screen off an entire lab and then treat it quantum mechanically. It is simply the case that the screened-off lab cannot interact with the universe, since that creates the possibility of inconsistent, shared classical information in the universe which constitutes a scientific contradiction. Therefore, the exchanges needed in FR to create the inconsistencies are the cause of the inconsistencies. This is where our constraint-based explanation would stand, since all classical information is to be represented in the universe in self-consistent fashion. But, there is no collapse on our view nor any question about how or when or why collapse happens. As we said at the beginning, we construct a realistic psi-epistemic account of QM \cite{stuckeyIJQF}.  
 
As we suggested in our initial discussion about the relationship between the classical and the quantum, the presumably classical experimental setup (or the many analogs of that in a natural setting) cannot be reduced away and plays an absolutely essential role in explaining so-called quantum outcomes; that is, the experimental setup must be treated as classical in order to use QM. So again, for us the fundamental explanatory role goes to what we call an adynamical global constraint applied to the spatiotemporal distribution of outcomes. Because of our hunch about an adynamical global constraint being fundamental, we based our account on the path integral formalism and seek a realist account with a single history. The point is that for us, the very idea that something could be both truly screened-off from the rest of the universe and also meaningfully treated classically, makes no sense.

Suppose there exists a ``quantum entity'' with some set of causal properties traversing the space between the source and detector to mediate a quantum exchange. According to environmental decoherence, this quantum entity cannot interact with its environment or it will cease to behave quantum mechanically, for example, it will act as a particle instead of a wave and cease to contribute to interference patterns. Now if this quantum entity does not interact in any way with anything in the universe (it is screened off), then it is not exchanging bosons of any sort with any object in the universe of classical physics. Thus, it does not affect the spacetime geometry along any hypothetical worldline (except at the source and detector) according to Einstein's equations (or it would be interacting gravitationally, i.e., exchanging gravitons). Accordingly, there cannot be any stress-energy tensor associated with the worldline of this quantum entity any place except at the source and detector. So, practically speaking, this posited ``screened-off quantum entity'' is equivalent to ``direct action.'' Thus, according to our view, QM is simply providing a probability amplitude for the spatiotemporal distribution of outcomes in the QM experiment. That is, in our view there are no ``QM systems'' such as waves, particles, or fields that exist independently of spatiotemporal, classical contexts. As Feynman puts it \cite[p. xv]{feynmanPI}:
\begin{quote}
In the customary view, things are discussed as a function of time in very great detail. For example, you have the field at this moment, a different equation gives you the field at a later moment and so on; a method, which we shall call the Hamiltonian method. We have, instead [the action] a thing that describes the character of the path throughout all of space and time . . . . From the overall space-time point of view of the least action principle, the field disappears as nothing but bookkeeping variables insisted on by the Hamiltonian method.
\end{quote}
Given our realist psi-epistemic account of QM with unmediated exchanges/direct action (i.e., there are no worldlines of counterfactual definiteness that connect the source and the detector), there is no ``screened off quantum entity'' that must decohere to behave classically. So on our view, environmental/dynamical decoherence is really just a dynamical take on what is in fact spatiotemporal contextuality. What QM is really telling us is that to exist (i.e., to be a diachronic entity in space and time), is to interact with the rest of the universe creating a consistent, shared set of classical information constituting the universe. 

Consistent with this quantum-classical ontology is the distinction between quantum and classical statistics, as characterized by the Born rule. That is, one adds individual amplitudes then squares to get the probability, rather than squaring each amplitude then adding as in classical probability, e.g., statistical mechanics. The quantum exchange of energy-momentum between classical objects requires cancellation of possibilities a la the path integral whereby the spacetime path of extremal action (classical trajectory) is obtained by interference of non-extremal possibilities which contribute with equal weight \cite[p. 224-225]{shankar}. Classical statistics does not provide for this so-called quantum interference. In fact, on our view the reason classical statistics works for classical objects is precisely because a classical object is a set of definite (high probability) quantum exchanges in the context of all other classical objects, as we just explained. That is, classical statistics assumes a distribution of classical objects and any given classical object can be viewed as obtaining from quantum statistics in the context of all other classical objects, having removed non-extremal possibilities via quantum interference. Therefore, classical statistics follows from quantum statistics as the quantum exchange of energy-momentum must be in accord with the classical objects of the universe. Again, since our model of objective reality is based on spatiotemporal-global self-consistency, the quantum/classical is not more fundamental than the classical/quantum -- they require each other. Notice again, anything, any piece of equipment, an elephant, etc., could in principle be screened-off and treated quantum mechanically. But again, the thing in question could only be so analyzed in some classical context. TTOs are defined relationally via their interactions with other TTOs, so no interactions means no TTO. This means the entire collection of TTOs cannot be decomposed quantum mechanically ``at once.'' The so-called ``quantum system'' is in fact the totality of the entire experimental set-up \cite[p. 738]{gomatam}, such that different set-ups or configurations are not probing some autonomous quantum realm, but actually constitute different ``systems.'' As with all the other mysteries of QM, the Born rule itself is vexing only if one assumes there is an ontology (created from quantum information) fundamental to that of the classical objects of the universe. Moving to the spacetime perspective allows one to consider an entirely new fundamental ontology, a quantum-classical ontology based on an adynamical global constraint. Indeed, per the Feynman path integral for QM, the most probable path is the classical path and per the transition amplitude for quantum field theory, the most probable field configuration is the classical field configuration \cite[Chap. 5]{ourbook}.
 
In the Bohmian account (a relative-state formalism) of FR, Lazarovici \& Hubert write \cite{lazarovici}:
\begin{quote}
the macroscopic quantum measurements performed by [Zeus] and [Wigner] are so invasive that they can change the actual state of the respective laboratory, including the records and memories (brain states) of the experimentalists in it.
\end{quote}
Per Lazarovici \& Hubert, memories and records change, but the history of those memories and records (along their worldlines prior to measurement) remain intact, so nothing in the past is changed. It is exactly analogous to passing vertically polarized light through a polarizer at $45^0$ then measuring it horizontally. The light incident on the first polarizer at $45^0$ has no horizontal component, but it does after passing through the polarizer at $45^0$. Consequently, it can now pass through the horizontal polarizer. Thus, for the photons that are now passing through the horizontal polarizer, the polarizer at $45^0$ can be said to have changed them from vertically polarized to horizontally polarized. Likewise, Zeus and Wigner's $\hat{z}$ and $\hat{w}$ measurements can literally change Xena and Yvonne's records and memories of their $\hat{x}$ and $\hat{y}$ measurement outcomes. 

This does not necessarily constitute scientific contradiction. If Xena and Yvonne's classical information prior to being measured by Zeus and Wigner is not shared (so that their worldlines are not part of the block universe of objective reality), and the classical information that exists at the end of the experiment that is shared by all participants is not self-contradictory, then there is no scientific contradiction. Of course, it may be impossible to tell a self-consistent dynamical story about how the initial self-consistent set of classical information evolved into the final self-consistent collection of classical information, but again that is not a problem for adynamical explanation. 

Healey also formulates a relative-states approach to FR, which we might infer from his statement \cite{healeyIJQF}:
\begin{quote}
So one could argue that whatever Wigner says about his outcome (more carefully, whatever Zeus measures Wigner’s outcome to be) is not a reliable guide to Wigner's actual outcome. In particular, even if Zeus takes Wigner's outcome to have been OK (because that’s what he observes it to be in a hypothetical future measurement on W) Wigner's actual outcome might equally well have been FAIL. That is, Zeus and Wigner's outcomes for measurements on Xena and Yvonne's records can contradict those records.
\end{quote}
This violates BW and FR's assumption of consistency only if you subscribe to the belief that QM probabilities apply to an objective reality. Healey and other relative-state approaches simply deny that assumption. Per Healey's pragmatic account of QM \cite{healeybook}, the job of QM is simply to provide the probabilities/correlations for outcomes in a quantum experiment given the experimental context for each observer, i.e., QM is not providing a physical model or interpretation of what happens between experimental initiation and termination -- in Bub's wording, ``the non-Boolean link'' between the Boolean initial conditions and the Boolean outcomes. Whereas our constraint-based account of QM provides a physical model that includes direct action, Healey's pragmatic account embraces metaphysical quietism about what happens between the initiation and termination of an experimental set-up. Hence his denial that QM probabilities describe an objective reality.  

Assuming the existence of a true quantum system $|\Psi\rangle$ represented by Eq. (\ref{Eq13}) and the standard formalism, Wigner and Zeus share a common classical context for making their measurements of $|\Psi\rangle$. In the relative-state formalism, Wigner/Zeus must treat Zeus/Wigner as a third quantum system resulting in a new version of Eq. (\ref{Eq13}) if Zeus/Wigner makes his measurement first. In the standard formalism, Eq. (\ref{Eq13}) is used by both Zeus and Wigner to determine distributions in their common spacetime context for whatever measurements they decide to make, since their measurements act on different parts of $|\Psi\rangle$ (Zeus on Xena's lab and Wigner on Yvonne's lab). Thus, the order of their measurements does not affect the predicted probabilities. For example, per the standard formalism, regardless of what Zeus measures, the probability that Wigner will get an OK outcome for a $\hat{w}$ measurement if Xena got tails in her $\hat{x}$ measurement is zero. That is because the tails part of Eq. (\ref{Eq13}) is

\begin{equation} 
\frac{1}{\sqrt{3}}| tails \rangle |fail \rangle_W\label{Bub3}
\end{equation}
But, in the relative-state formalism, this same probability depends on whether or not Zeus makes his measurement first and what measurement Zeus makes, since the functional form of $|\Psi\rangle$ will be different for Wigner if Zeus makes an intervening measurement.

For example, suppose Zeus measures $\hat{x}$ first. After Zeus's measurement per the relative-state formalism, Eq. (\ref{Eq13}) reads

\begin{equation} 
|\Psi \rangle = \frac{1}{\sqrt{3}}\left(| heads \rangle_X | heads \rangle_Z | - \rangle + | tails \rangle_X | tails \rangle_Z |fail \rangle_W \right)\label{Bub4}
\end{equation}
[Notice we must now distinguish Xena from Zeus even though they are measuring the same thing. The same must be done with Yvonne and Wigner.] In this case, as in the standard formalism, the probability that Wigner will get an OK outcome for a $\hat{w}$ measurement if Xena got tails in her $\hat{x}$ measurement is zero because the tails part of Eq. (\ref{Bub4}) is

\begin{equation} 
\frac{1}{\sqrt{3}}| tails \rangle_X | tails \rangle_Z |fail \rangle_W\label{Bub5}
\end{equation}
after Zeus's $\hat{x}$ measurement. But, suppose Zeus makes a $\hat{z}$ measurement instead. To use the relative-state formalism, we must first cast Eq. (\ref{Eq13}) in the OK-fail basis as \cite{bubforthcoming}

\begin{equation} 
|\Psi \rangle = \frac{1}{\sqrt{12}}| OK \rangle_X | OK \rangle_Y - \frac{1}{\sqrt{12}}| OK \rangle_X | fail \rangle_Y + \frac{1}{\sqrt{12}}| fail \rangle_X | OK \rangle_Y + \frac{\sqrt{3}}{2}| fail \rangle_X | fail \rangle_Y\label{BubEq16}
\end{equation}
Now, after Zeus makes his $\hat{z}$ measurement, Eq. (\ref{BubEq16}) reads \cite{bubforthcoming}

\begin{equation}
\begin{split}
|\Psi \rangle = &\frac{1}{\sqrt{12}}| OK \rangle_X | OK \rangle_Z | OK \rangle_Y - \frac{1}{\sqrt{12}}| OK \rangle_X | OK \rangle_Z | fail \rangle_Y +\\& \frac{1}{\sqrt{12}}| fail \rangle_X | fail \rangle_Z | OK \rangle_Y +\frac{\sqrt{3}}{2}| fail \rangle_X | fail \rangle_Z | fail \rangle_Y
\end{split}
\label{BubEq17}
\end{equation}
Thus, the tails part of Eq. (\ref{BubEq17}) is \cite{bubforthcoming}

\begin{equation}
|tails \rangle_X\left[ \frac{\sqrt{5}}{\sqrt{12}}\left(\frac{3}{\sqrt{10}}| fail \rangle_Z + \frac{1}{\sqrt{10}}| OK \rangle_Z \right)|fail \rangle_Y + \frac{1}{\sqrt{12}}\left(\frac{1}{\sqrt{2}}| fail \rangle_Z - \frac{1}{\sqrt{2}} | OK \rangle_Z \right) | OK \rangle_Y\right]
\label{BubEq18}\end{equation}
And since

\begin{equation}
\left(\frac{1}{\sqrt{2}}| fail \rangle_Z - \frac{1}{\sqrt{2}} | OK \rangle_Z \right) | OK \rangle_Y = |tails\rangle_Z | OK \rangle_Y
\label{BubEq20}\end{equation}
we now have a non-zero probability for Wigner obtaining an OK outcome for a $\hat{w}$ measurement when Xena obtains a tails outcome for her $\hat{x}$ measurement (it is $\frac{1}{6}$ actually \cite{bubforthcoming}). If Zeus does not make a measurement, Eq. (\ref{BubEq16}) becomes

\begin{equation}
\begin{split}
|\Psi \rangle = &\frac{1}{\sqrt{12}}| OK \rangle_X | OK \rangle_Y | OK \rangle_W - \frac{1}{\sqrt{12}}| OK \rangle_X | fail \rangle_Y | fail \rangle_W +\\& \frac{1}{\sqrt{12}}| fail \rangle_X | OK \rangle_Y | OK \rangle_W + \frac{\sqrt{3}}{2}| fail \rangle_X | fail \rangle_Y | fail \rangle_W
\end{split}
\label{BubEqX}\end{equation}
after Wigner's $\hat{w}$ measurement. The $|OK\rangle_W$ part of this is

\begin{equation}
\frac{1}{\sqrt{12}}\left(| OK \rangle_X + | fail \rangle_X \right) | OK \rangle_Y | OK \rangle_W = \frac{1}{\sqrt{6}} | heads \rangle_X | OK \rangle_Y | OK \rangle_W
\label{BubEqXX}\end{equation}
which has no tails piece for Xena, so the probability of Wigner obtaining an OK outcome for a $\hat{w}$ measurement when Zeus has not made a measurement and when Xena obtains a tails outcome for her $\hat{x}$ measurement is again zero.

You can see why Zeus's intervening $\hat{z}$ measurement per the relative-state formalism can cause possible contradictions between records (as in Healey's version of the relative-state formalism) or changes to records and memories (as in Lazarovici \& Hubert's version of the relative-state formalism). In the relative-state formalism, Zeus's measurement outcome must match Xena's hypothetical or recorded measurement outcome in the basis used by Zeus, e.g., $| OK \rangle_X | OK \rangle_Z$. Everything is fine as long as Zeus and Xena make the same measurement, but if Zeus measures in a rotated Hilbert space basis relative to Xena, his possible measurement outcomes will contain cross terms in Xena's possible measurement outcomes, e.g., $| heads \rangle_X | tails \rangle_Z$ and $| tails \rangle_X | heads \rangle_Z$, which implies a contradiction between what Zeus measures for Xena's measurement outcomes and what Xena actually measured and recorded. So that contradiction stands (as in Healey) or Xena's outcomes change (as in Lazarovici \& Huber). Neither of these radical responses is required in our single, self-consistent, spatiotemporal model of objective reality.


\section{\label{WFexperiment}Experimental Evidence for Wigner's Friend?}
\begin{figure}
\begin{center}
\includegraphics [height = 60mm]{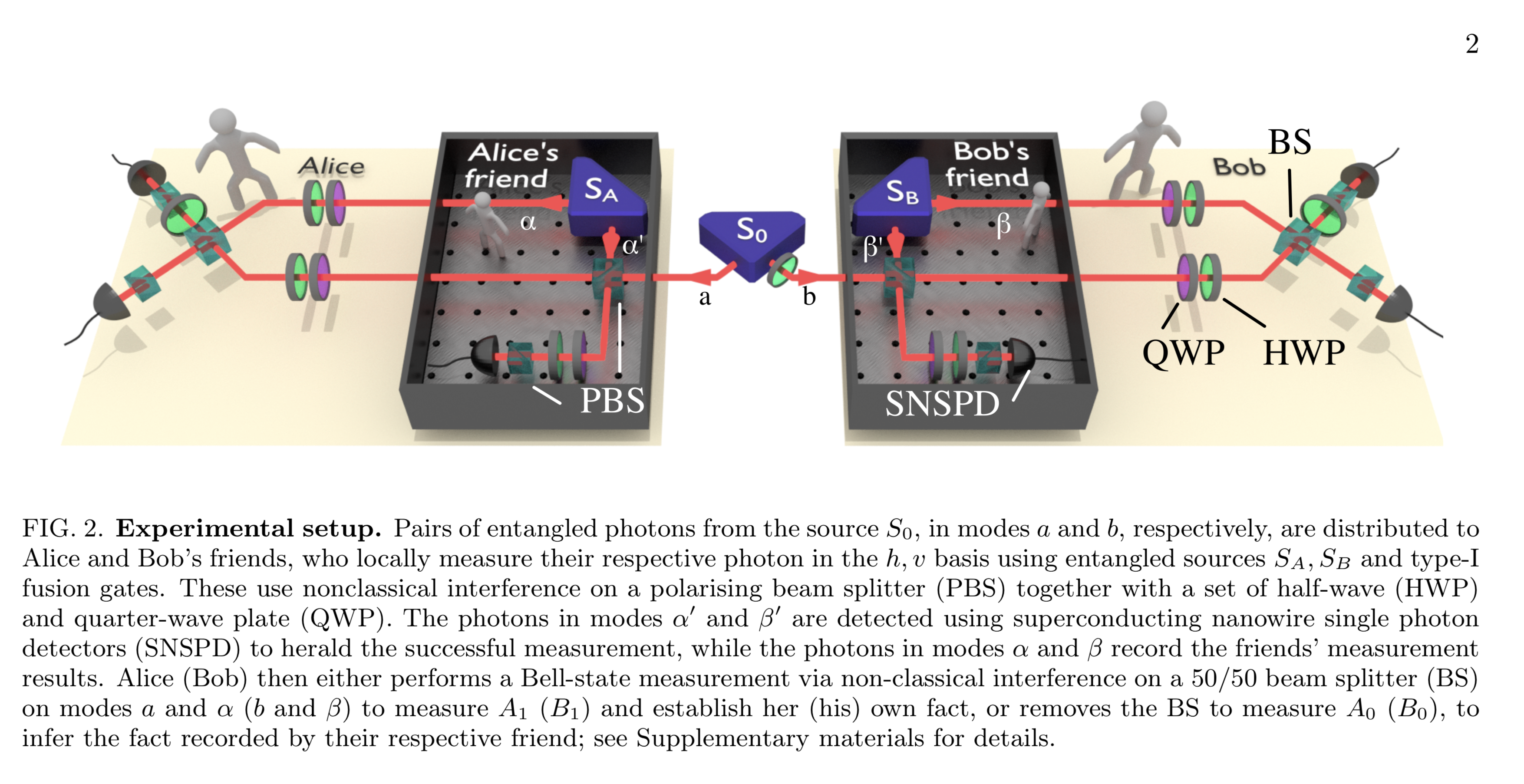}  \caption{Figure from Proietti et al. \cite{proietti} showing and explaining their experimental set-up.} \label{WFexp}
\end{center}
\end{figure}

In ``Experimental test of local observer-independence,'' Proietti et al. \cite{proietti} claim to have an experimental result which ``lends considerable strength to interpretations of quantum theory already set in an observer-dependent framework and demands for revision of those which are not.'' As their depiction of their experiment (Figure \ref{WFexp}) clearly shows, all the experimental measurements and outcomes for their experiment occur in a single objective reality, i.e., in the self-consistent, shared classical information of the spacetime model of objective reality. And, as they show in their paper, all these outcomes are in accord with QM. Thus, as Carroll notes  \cite{dailynous}:
\begin{quote}
What they have not done is to call into question the existence of an objective reality. Such a reality may or may not exist (I think it does), but experiments that return results compatible with the standard predictions of quantum mechanics cannot possibly overturn it.
\end{quote}
In Carroll's take on the experiment per Many-Worlds, Proietti et al. did not cause a branching since \cite{dailynous}:
\begin{quote}
Rather than having an actual human friend who observes the photon polarization -- which would inevitably lead to decoherence and branching, because humans are gigantic macroscopic objects who can't help but interact with the environment around them -- the ``observer'' in this case is just a single photon. For an Everettian, this means that there is still just one branch of the wave function all along. The idea that ``the observer sees a definite outcome'' is replaced by ``one photon becomes entangled with another photon,'' which is a perfectly reversible process. Reality, which to an Everettian is isomorphic to a wave function, remains perfectly intact.
\end{quote}
Maudlin agrees, saying  \cite{dailynous}:
\begin{quote}
The experiments in question are done on a system composed of only six photons. Obviously the photons do not experience anything at all, much less conflicting realities.
\end{quote}
And Crowther says \cite{dailynous}:
\begin{quote}
But, on the other hand, the fact remains that these devices are not conscious, and so Wigner could stand resolute in his interpretation. If anything, he could point out that -- in the same way that an observation of a non-black, non-raven provides a negligible sliver of confirmation for the claim that `all ravens are black' -- the success of the experiment even provides inductive support in favour of his interpretation: the `observers' in this experiment are able to record conflicting facts only because they do not experience these facts. 
\end{quote}
In other words, Proietti et al. did not in any way screen off a macroscopic measurement device and outcome recording -- all measurement devices are visible in Figure \ref{WFexp} and all the experimental outcomes at all times are accessible to all observers in spacetime. Thus, the experiment constitutes a self-consistent (per QM) collection of observations in the spacetime model of objective reality. Indeed, again, one cannot even employ the formalism of QM without coordination of Cartesian frames and synchronization of clocks throughout the spacetime region of the experiment. As Lazarovici notes  \cite{dailynous}:
\begin{quote}
A group of physicists claims to have found experimental evidence that there are no objective facts observed in quantum experiments. For some reason, they have still chosen to share the observations from their quantum experiment with the outside world.
\end{quote}
Maudlin agrees on this point as well, saying  \cite{dailynous}:
\begin{quote}
If there is no objective physical world then there is no subject matter for physics, and no resources to account for the outcomes of experiments.
\end{quote}
Thus, the Proietti et al. experiment neither establishes the claim in the title of their paper nor provides a true instantiation of the Wigner's friend experiment. However, we believe their experiment does contain an interesting hint of what we pointed out earlier is otherwise ignored in Wigner's friend scenarios. 

That is, their experiment does hint at what it might mean for Wigner to measure his friend's lab and measurement results in a rotated Hilbert space basis. Wigner's non-rotated Hilbert space basis (the direct measurement of Wigner's friend's result) is achieved in Proietti et al. by removing the beam splitter(s) (BS) in Figure \ref{WFexp} thereby measuring $A_o$ and/or $B_o$. That means Wigner (here represented by Alice and Bob) is measuring directly both the friend's ``measurement system'' (lower exiting red beam on either side) and the friend's ``outcome recording'' (upper exiting red beam on either side). Inserting the beam splitter(s) then mixes this information, as represented by a rotated Hilbert space basis. 

Of course, in such a case Wigner's friend's result would not be in conflict with Wigner's measurement for two reasons. First, Wigner's result is a quantum conflated measurement of his friend's outcome and measurement, so there could be no contradiction in such a result even if there was some way to make sense of the friend's measurement and outcome when screened off. That is the case in Proietti et al., since the entire experimental set-up exists in the spacetime model of shared classical information. Second, as we stated earlier, when Wigner's friend and his lab are screened off from the rest of the universe (contrary to Proietti et al.) there is no common classical context in which to interpret the friend's measurement and outcome. So, it would be impossible for any contradiction to be observed, i.e., there is no violation of the consistency of shared classical information constituting spacetime.  

Simply put, the results that violate Bell's inequality in the Proietti et al. experiment imply, at worst, that there is no counterfactual definiteness for screened-off quantum systems contributing to the spacetime of shared classical information, just as in any other violation of Bell's inequality. The experiment Proietti et al. should have claimed to instantiate is the quantum liar experiment of Elitzur \& Dolev \cite{avi,QLE}. 

In the quantum liar experiment, an experimental configuration leads to the creation of a quantum state which then violates the Bell inequality. But, the violation of the Bell inequality by this state denies the very counterfactual definiteness responsible for creating the state to begin with. Again, this does not violate the consistency of shared classical information constituting the spacetime model of objective reality \cite{QLE}. In Proietti et al. the Bell-inequality-violating states represent quantum information about a measurement and its outcome. As with any other combined quantum information, quantum interference can then erase various individual contributions. This interference does not change the friend's measurement and outcome, it just changes the information concerning the friend's measurement and outcome, and it does so without jeopardizing the self-consistency of shared classical information constituting the spacetime model of objective reality. This is precisely what happened in Proietti et al.

In ``A strong no-go theorem on the Wigner's friend paradox,'' Bong et al. also claim to have a ``proof-of-principle'' experiment for such ``extended Wigner's friend scenarios'' (EWFS) \cite{Bong2020}. Specifically, as we said at the beginning, they derive an inequality that when violated by any physical theory entails the violation of at least one of the following assumptions in the context of EWFS (worded colloquially here by Cavalcanti) \cite{Cavalcanti2020}:
\begin{enumerate}
    \item When someone observes an event happening, it really happened.
    \item It is possible to make free choices, or at least, statistically random choices.
    \item A choice made in one place can't instantly affect a distant event.
\end{enumerate}
They refer to these assumptions collectively as Local Friendliness (LF). Their LF theorem is:
\begin{quote}
    If a superobserver [Wigner or Zeus above] can perform arbitrary quantum operations on an observer and its environment [Xena or Yvonne above], then no physical theory can satisfy Local Friendliness.
\end{quote}
Assumption 1 is the ``Absoluteness of Observed Events (AOE)): An observed event is a real single event, and not relative to anything or anyone.'' As they point out, this is a tacit assumption made in deriving Bell inequalities, e.g., the CHSH inequality. And as we showed above, this assumption is necessary for using the formalism of QM, i.e., if one violates this assumption when using the formalism of QM, contradictions and absurdities can arise. So, while it may certainly be true that violating Assumption 1 leads to the violation of their LF inequality, one cannot use QM to check their LF inequality while violating Assumption 1. That means a violation of their LF inequality by QM entails the violation of either or both of ``Assumption 2 (No-Superdeterminism (NSD)): Any set of events on a space-like hypersurface is uncorrelated with any set of freely chosen actions subsequent to that space-like hypersurface'' or ``Assumption 3 (Locality (L)): The probability of an observable event \textit{e} is unchanged by conditioning on a space-like-separated free choice \textit{z}, even if it is already conditioned on other events not in the future light-cone of \textit{z}.'' Of course, NSD and L are just the assumptions for the Bell inequality, so other than an new inequality nothing new has been introduced in this paper regarding QM. 

As with Proietti et al., the spacetime region for the Bong et al. experiment resides entirely in the spacetime of self-consistent, shared classical information, so we can use QM to check their experiment for the violation of their LF inequality. Again, if the LF inequality is violated by QM (as it is for this experiment), we only know that we must abandon NSD or L, since AOE is required to map the QM formalism to experimental arrangements. In this experiment, the ``superobservers'' are Alice and Bob and their friends are Charlie and Debbie, respectively. No measurement outcomes for Charlie and Debbie are involved in the LF inequality, so we may dismiss them immediately; again, this is certainly not an EWFS (which they readily admit in the paper). Their source produces a mixture that they can tune via $\mu$
\begin{equation}
    \rho_\mu = \mu |\Phi^- \rangle \langle \Phi^-| + \frac{1-\mu}{2} (|HV\rangle \langle HV| + |VH\rangle \langle VH| )
\end{equation}
where $|\Phi^- \rangle$ is the spin singlet state. They find the maximum violations of the Bell and LF inequalities occur when the state is tuned entirely to the spin singlet state ($\mu = 1$), so we'll briefly review that case.

\begin{figure}
\begin{center}
\includegraphics [height = 30mm]{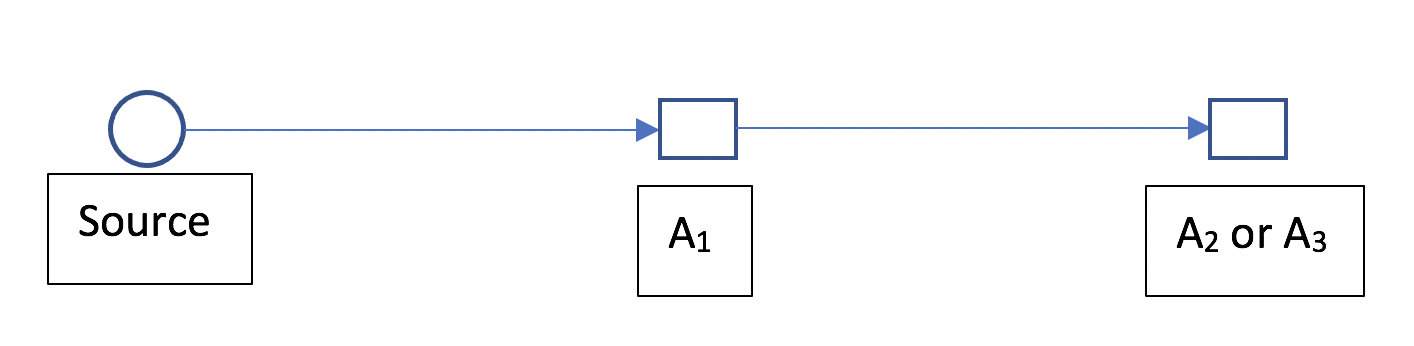}  \caption{Alice's side of Bong et al. experiment. Bob's side is the same with $B_j$ replacing the $A_i$.} \label{BongExp1}
\end{center}
\end{figure}

\begin{figure}
\begin{center}
\includegraphics [height = 60mm]{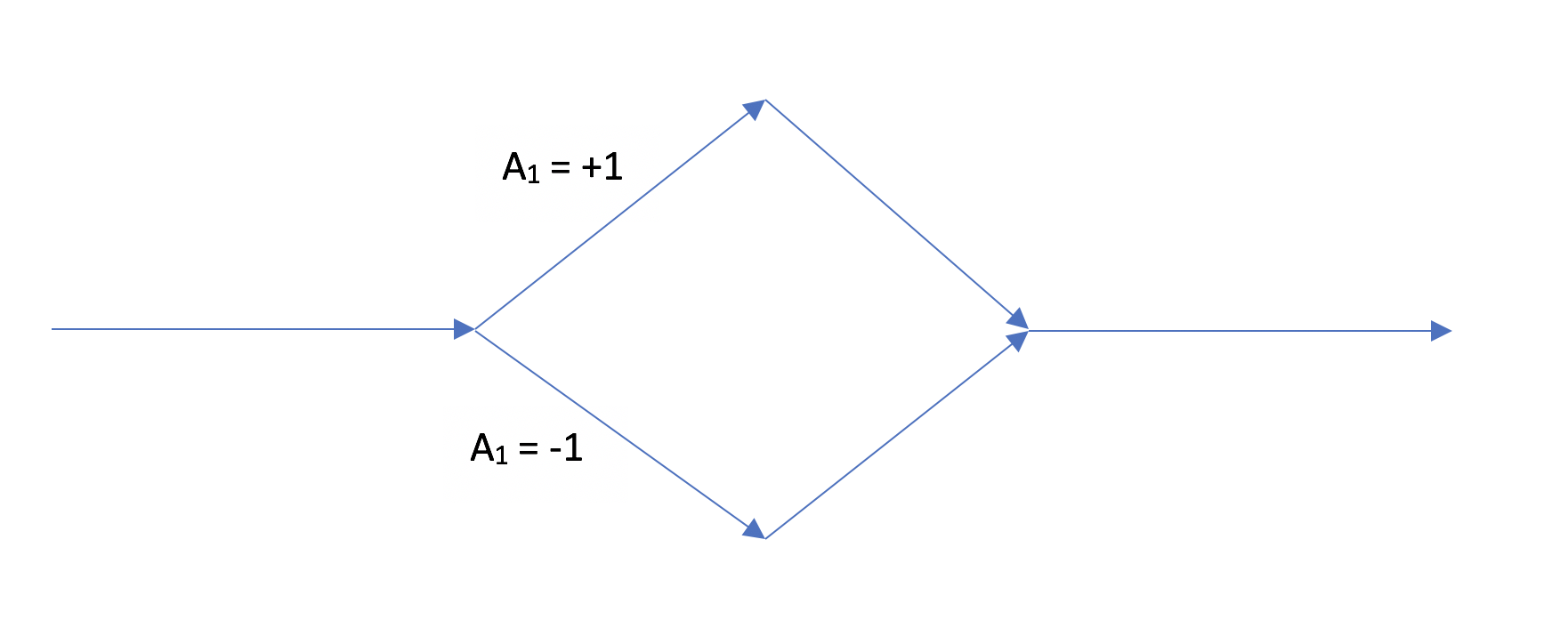}  \caption{If the measurement $A_1$ is done, the particle is intercepted on the upper path ($A_1 = +1$) or the lower path ($A_1 = -1$). Otherwise, the particle is allowed to continue to either an $A_2$ or $A_3$ measurement. In order to promote this to a ``proof-of-principle'' EWFS experiment, the path inside here is taken to represent Charlie (Debbie on Bob's side) having made his(her) measurement and obtained his(her) result.} \label{BongExp2}
\end{center}
\end{figure}

\begin{figure}
\begin{center}
\includegraphics [height = 60mm]{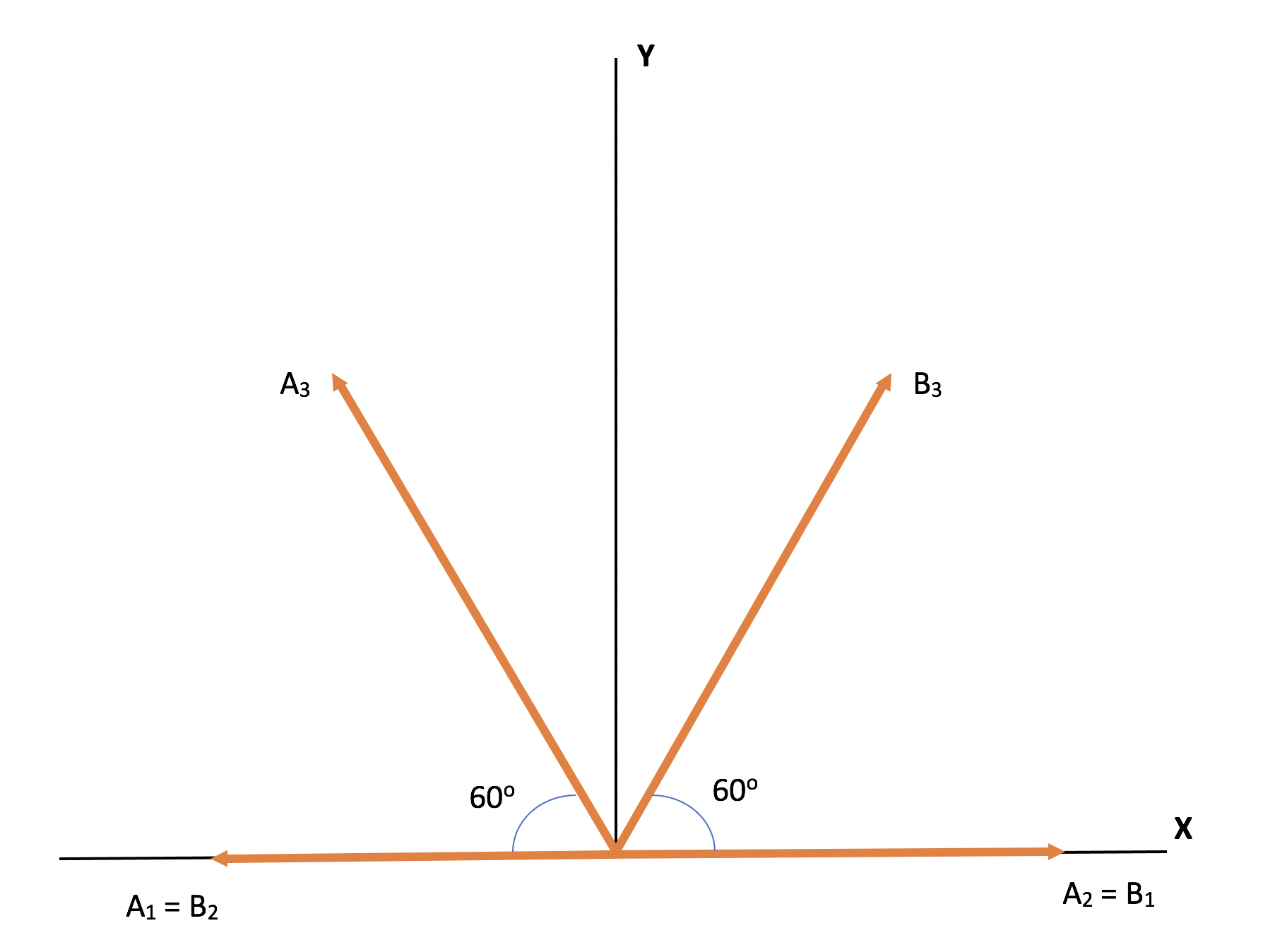}  \caption{Spin measurements $A_i$ and $B_j$ in the $xy$ plane.} \label{SpinMeasurements}
\end{center}
\end{figure}

Alice and Bob choose between three spin measurements labeled $\{A_1,A_2,A_3\}$ and $\{B_1,B_2,B_3\}$, respectively, in each trial of the experiment as shown schematically in Figures \ref{BongExp1} \& \ref{BongExp2}. In the context of these measurements, the LF inequality is
\begin{equation}
-\langle A_1 \rangle - \langle A_2 \rangle - \langle B_1 \rangle - \langle B_2 \rangle - \langle A_1B_1 \rangle - 2\langle A_1B_2 \rangle - 2\langle A_2B_1 \rangle + 2\langle A_2B_2 \rangle - \langle A_2B_3 \rangle - \langle A_3B_2 \rangle - \langle A_3B_3 \rangle -6 \le 0
\end{equation}
By comparison, the CHSH (Bell) inequality for these measurements is
\begin{equation}
\langle A_2B_2 \rangle - \langle A_2B_3 \rangle - \langle A_3B_2 \rangle - \langle A_3B_3 \rangle -2 \le 0
\end{equation}
These inequalities are saturated for the counterfactually definite set $\{A_1,A_2,A_3,B_1,B_2,B_3\} = \{1,-1,1,-1,-1,-1\}$, for example. In order to evaluate them for the spin singlet state, we choose the measurements shown in Figure \ref{SpinMeasurements}. These choices are close enough to those used by Bong et al. so as to reproduce their results within experimental error. For the spin singlet state we have $\langle A_i \rangle = \langle B_j \rangle = 0$ and $\langle A_iB_j \rangle = -\cos\theta$ where $\theta$ is the angle between the measurements $A_i$ and $B_j$ as shown in Figure \ref{SpinMeasurements} \cite{MerminChallenge}. With these measurements of the spin singlet state the lefthand sides of the LF inequality and the CHSH (Bell) inequality are both 0.5, in agreement with the Bong et al. results. 

Essentially, the Bong et al. experiment is just another version of the Greenberger-Horne-Zeilinger (GHZ) experiment \cite{GHZ}, which Dowker uses to motivate Sorkin's Many Histories interpretation of QM \cite{Dowker2014}. That is, the violation of the CHSH (Bell) inequality tells against counterfactual definiteness (no ``instruction sets'' per Mermin \cite{merminGHZ}) which entails no definite path for the photon through the experimental arrangement unless explicitly measured, e.g., $A_1$ ($B_1$). But, if the photon's path through region $A_1$ ($B_1$) is not definite under measurements $A_2$ or $A_3$ ($B_2$ or $B_3$), then what does that imply about Charlie or Debbie's measurement outcomes? You can see how this would bear on the mystery of the EWFS, if in fact they had actually screened off a classical measurement and recording device, but they did not, so they're left with another (clever) quantum entanglement experiment. 

\section{\label{delayed}Principle Explanation of Delayed Choice Quantum Eraser}
In addition to Wigner's friend, another obvious case where there is a possible tension between how we experience the world and some QM experiment is delayed choice quantum eraser. So, in this section, we consider constraint-based explanation for the delayed choice quantum eraser experiment. In his chapter in this volume Hardy will outline a possible real experiment of the sort we discuss herein, whereas our version is but a toy experiment for the purposes of illustration. Thus, in order to bring this possible tension out most fully we will alter the set-up of the experiment by adding a conscious agent who attempts to violate the probabilities of QM, as one might think a truly free conscious agent ought to be able to do. Let us start with a description of the experiment.

\begin{figure}
\begin{center}
\includegraphics [height = 60mm]{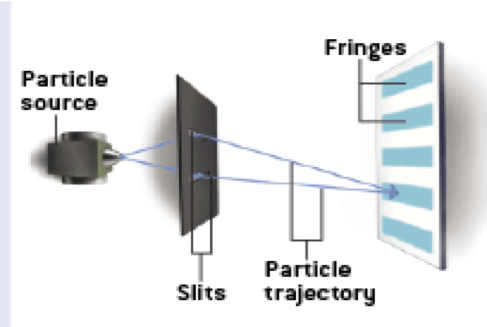}  \caption{Particles create an interference pattern when proceeding through the double slits (figure from Hillmer and Kwiat \cite{hillmer}).} \label{DC1}
\end{center}
\end{figure}

\begin{figure}
\begin{center}
\includegraphics [height = 70mm]{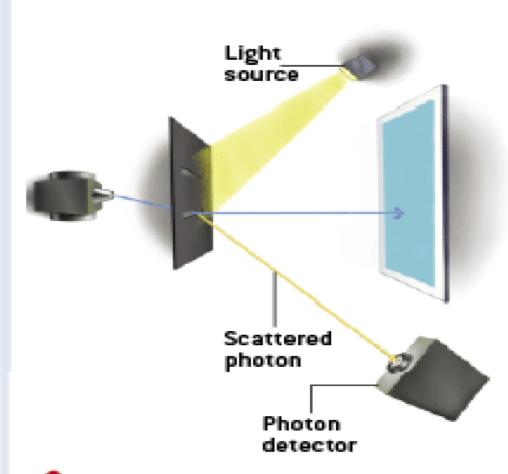}  \caption{The interference pattern of Figure \ref{DC1} can be destroyed by scattering photons and using those scattered photons to determine which slit the particle went through on each trial (figure from Hillmer and Kwiat \cite{hillmer}).} \label{DC2}
\end{center}
\end{figure}

\begin{figure}
\begin{center}
\includegraphics [height = 70mm]{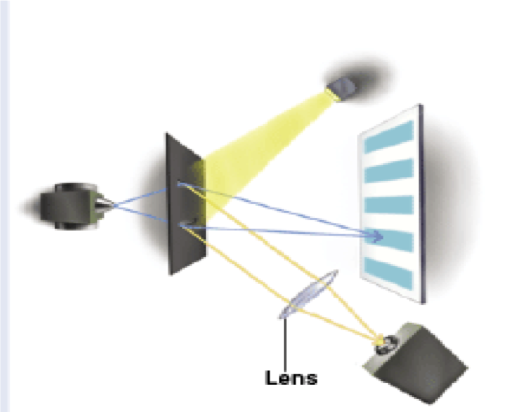}  \caption{The interference pattern of Figure \ref{DC1} can be restored after scattering photons as in Figure \ref{DC2} by destroying the which-way information in the scattered photons (here done by inserting a lens). This is known as ``quantum eraser'' (figure from Hillmer and Kwiat \cite{hillmer}).} \label{DC3}
\end{center}
\end{figure}

Using pictures from Hillmer and Kwiat \cite{hillmer} we start with a particle interference pattern (Figure \ref{DC1}) then we scatter photons off the particles after they have passed through the slits(s) (Figure \ref{DC2}) and finally we erase the which-way information obtained by the scattered photons by inserting a lens (Figure \ref{DC3}). 

In the Hillmer and Kwiat article the lens (eraser) is inserted after the particles have passed through the slits, but experiments have been done where the `lens is inserted' after the particles have hit the detector. This is called a ``delayed choice quantum eraser experiment'' \cite{kim}. The question from our dynamical perspective is, How do the particles `know' whether or not the lens will be inserted? And, if they do not `know' whether the lens will be inserted or not, how do they `know' whether or not to create the interference pattern? These questions assume temporally sequential, causal explanation, i.e., we are playing chess. 

If we rather seek an adynamical, spatiotemporal constraint-based explanation in crossword puzzle fashion, we are content with the fact per QM that the distribution of particles on the screen is consistent with the presence or absence of the lens in spacetime. The insertion of the lens does not `cause' the interference pattern any more than the interference pattern `causes' the insertion of the lens. No new physics is needed to explain this phenomenon, just the willingness to rise to Wilczek's challenge \cite[p. 37]{wilczek}:
\begin{quote}
A recurring theme in natural philosophy is the tension between the God's-eye view of reality comprehended as a whole and the ant's-eye view of human consciousness, which senses a succession of events in time. Since the days of Isaac Newton, the ant's-eye view has dominated fundamental physics. We divide our description of the world into dynamical laws that, paradoxically, exist outside of time according to some, and initial conditions on which those laws act. The dynamical laws do not determine which initial conditions describe reality. That division has been enormously useful and successful pragmatically, but it leaves us far short of a full scientific account of the world as we know it. The account it gives -- things are what they are because they were what they were -- raises the question, Why were things that way and not any other? The God's-eye view seems, in the light of relativity theory, to be far more natural. ... \textit{To me, ascending from the ant's-eye view to the God's-eye view of physical reality is the most profound challenge for fundamental physics in the next 100 years} [italics ours].
\end{quote}
Let us now bring the conscious agent into the picture by imagining it is a conscious agent inserting the lens (or not) in the experimental set-up. The question from our dynamical perspective is, What will I experience if I am the agent deciding whether or not to insert the lens? If the predictions of QM are to hold, then my decision must always be in accord with the particle's behavior at the detection screen and that event occurred before I made my decision. Assuming QM holds, will I feel mentally `coerced' into making the appropriate choice? Will I feel some `physical force' moving my hand against my will? Most people do not like the idea that our ``freely made'' decisions can be the result of a single particle striking a distant detector. It would seem that QM does not care about choice at all, delayed or otherwise. 

While most people predict that a conscious agent will not violate the probabilities of QM anymore than a classical measuring device, Hardy has proposed an experiment to test this fact. Concerning such an experimental test, he states \cite{anan}:
\begin{quote}
[If] you only saw a violation of quantum theory when you had systems that might be regarded as conscious, humans or other animals, that would certainly be exciting. I can't imagine a more striking experimental result in physics than that. We'd want to debate as to what that meant. It wouldn't settle the question, but it would certainly have a strong bearing on the issue of free will.
\end{quote}
What explains the agreement between the agent's decision and the particle's pattern if it is not ``spooky action at a distance'' or ``backwards causation?'' Why does the conscious agent always (statistically at least) make the ``right'' choice in accord with QM? One doubts there is some special new physical or mental force acting on the hand or mind of the conscious observer. For us the answer is simple -- we instead ignore our anthropocentric bias and allow for the possibility that objective reality is fundamentally the spacetime of shared, self-consistent classical information whose various patterns/distributions are determined fundamentally by adynamical global constraints, not by dynamical laws/processes acting on matter/mind to make it move/decide. We can then accept that there are some constraint-based explanations that do not allow for dynamical counterparts, at least dynamical counterparts without serious baggage, such as those discussed earlier. The constraint-based explanation here is the distribution of quantum energy-momentum exchanges in the spacetime context for the experimental set-up and procedure according to the adynamical global constraint of QM, as in section \ref{intro2}.

The point is, adynamical global constraints in spacetime also constrain the choices of conscious agents. Thus, physics is already part of psychology in that it places real constraints on what can be experienced to include memories (classical records) and choices. Conscious agents attempting to override QM do not experience any weird forces acting on them because there are no such forces. It is simply the case that their choices will be made in accord with the relevant adynamical global constraints per spacetime. Such agents feel like they have libertarian free will (that the future is open) because they experience reality from the ``ant's-eye'' view.   


\section{\label{Conclusion}Re-Thinking the World with Neutral Monism: Removing the Boundaries Between Mind, Matter, and Spacetime}

Let us now conclude by putting it all together, that is, our axiomatic principle constraints, our constraint-based explanation of the experiments herein, and neutral monism. How did we get to the point where every decade or two we feel compelled to try and relate the hard problem of consciousness, the measurement problem and mystery of free will? We get there by making certain assumptions. We believe it is high time we jettisoned these assumptions and start again with our best science as our guide. In particular we think the offending assumptions are: 1) physicalism, 2) fundamentalism, and relatedly 3) dualism about conscious experience, 4) the notion that fundamental explanation is always constructive, causal or dynamical, and relatedly, 5) realism about the wavefunction. Together these assumptions force us into the hard problem, they force us into the measurement problem, and they force us to seek the solutions in or add the solutions to fundamental physics, e.g., ``panpsychist fusion'' and all the rest \cite{NMEntropy2020}. Herein we have shown you an account of QM, relativity and their relationship to conscious observers that rejects all these assumptions. To appreciate fully how it all hangs together one must really appreciate neutral monism, so we will begin there.  

In its most general form neutral monism is the idea that mental and material features are real but in some specified sense, reducible to or constructable from a neutral basis in a non-eliminative sense of reduction. The neutral basis is not a substance. Mental and material features are not separable or merely correlated, they are non-dual, indeed, they are not essentially different and distinct aspects. Thus, experience isn’t inherently or essentially `inner' or mental and the `external’ world isn’t inherently non-mental. The particular brand of neutral monism we want to defend herein is most closely associated with William James and to a lesser degree Bertrand Russell. To hopefully help the reader to appreciate the idea here are some passages from James that we believe captures its character:
\begin{quote}
Subjectivity and objectivity are affairs not of what an experience is aboriginally made of, but of its classification \cite[p. 1208]{jamesN}.\\

``Subjects'' knowing,``things'' known, are ``roles'' played. Not ``ontological'' facts \cite[p. 110]{jamesPAF}. \\

The neutral ``in itself, is no more inner than outer ... . It becomes inner by belonging to an inner, it becomes outer by belonging to an outer, world'' \cite[p. 217]{jamesML}.\\

A given undivided portion of experience, taken in one context of associates, play[s] the part of the knower, or a state of mind, or ‘consciousness'; while in a different context the same undivided bit of experience plays the part of a thing known, of an objective ‘content.’ In a word, in one group it figures as a thought, in another group as a thing \cite[p. 533]{james1904b}.
\end{quote}
This idea can also be found in Hinduism and Buddhism long before it appears in the West. Take the following from Evan Thompson for example \cite[p. 61]{thompson2015}:
\begin{quote}
Take a moment of visual awareness such as seeing the blue sky on a crisp fall day. The ego consciousness makes the visual awareness feel as if it’s ‘my’ awareness and makes the blue sky seem the separate and independent object of ‘my’ awareness. In this way, the ego consciousness projects a subject–object structure onto awareness. According to the Yogacara philosophers, however, the blue sky isn’t really a separate and independent object that’s cognized by a separate and independent subject. Rather, there’s one ‘impression’ or ‘manifestation’ that has two sides or aspects—the outer-seeming aspect of the blue sky and the inner-seeming aspect of the visual awareness. What the ego consciousness does is to reify these two interdependent aspects into a separate subject and a separate object, but this is a cognitive distortion that falsifies the authentic character of the impression or manifestation as a phenomenal event. 
\end{quote}
Let us now relate this all back to physics. In neutral monism, what we call spacetime is nothing but the events therein and those events are neither inherently mental nor inherently physical. Russell calls such ``neutral'' events ``unstructured occurrences,'' such as ``hearing a tyre burst, or smelling a rotten egg, or feeling the coldness of a frog''  \cite[p. 287]{russell10}. How do physical phenomena relate to these neutral events? As Russell puts it, ``Matter and motion . . . are logical constructions using events as their material, and events are therefore something quite different from matter in motion'' \cite[p. 292]{russell10}. How we ultimately taxonomize those events, is as he says ``a mere linguistic convenience to regard a group of events as states of a `thing', or `substance', or `piece of matter', or a `precept' '' \cite[p. 284]{russell9}. Going further he says, ``electrons and protons . . . are not the stuff of the physical world'' \cite[p. 386]{russell10}. Again, ``bits of matter are not among the bricks out of which the world is built. The bricks are events, bits of matter are portions of the structure to which we find it convenient to give separate attention'' \cite[p. 329]{russell7}.

Therefore given neutral monism, the world is not made of or realized by essentially physical entities such as the QM wavefunction. Rather, what we call physical entities are contextually given manifestations of the neutral base. We believe this idea comports well with QM contextuality and relativity. Once one accepts such neutral monism and contextuality, it ought to lead one to question other things as follows. First, the notion of beables as hidden, distinct entities with metaphysical autonomy that are responsible for all observables is questionable. Per neutral monism, sometimes reality (i.e., spacetime) manifests as particle-like, field-like or wave-like, etc., depending on multiscale context, e.g., the twin-slit experiment. There are no context-independent beables, multiscale contextuality itself is fundamental. Second, constructive, constitutive, dynamical, and causal mechanical explanations are not always fundamental. Sometimes principle explanations a la spatiotemporal adynamical global constraints are fundamental, e.g., the light postulate, conservation laws, least action principles, etc. This isn’t surprising since the contextuality in question is spatiotemporal.   

What is the neutral base you ask? The neutral base in question is what James calls ``unqualified actuality'' and ``the instant field of the present.'' We call it Neutral Pure Presence (i.e., ``pure being'') or ``Nowness.'' Philosophers and physicists such as Einstein have long noted that there is something special about the experience of Nowness which is ``outside the realm of science,'' as he put it. Neutral monism holds that Presence is fundamental and universal. To paraphrase Hawking, it’s ``what puts the fire in the equations.'' 

This is not panpsychism. Panpsychism by definition is the view that whatever fundamental physical entities are, their intrinsic nature and essence is proto-qualia or proto-subjectivity. Given neutral monism, ``physical entities'' are manifestations of Presence and Presence is neutral, not mental. Unlike panpsychism which merely moves the mysterious dualism from brains to fundamental physical entities, neutral monism is a complete rejection of the primary/secondary property distinction. Panpsychism likes to fancy itself as a kind of dual-aspect monism, but that's just another name for property dualism. Panpsychism claims as an advantage over strong emergence that the origins of consciousness or proto-consciousness is in fundamental physics. The panpsychist thinks it should be comforting to us naturalists to associate consciousness with fundamental physics. We are deeply puzzled by this intuition. For us, panpsychism doesn’t make proto-qualia/proto-subjectivity any less weird, on the contrary. At least associating conscious minds with brains makes some intuitive and empirical sense; after all, there are many important dynamical and causal relationships between brain states and conscious states. Rather, panpsychism only makes matter weirder and seemingly less natural. It’s like learning that there are fairies in the world, but then being told to relax because we have decided they are just brute features of fundamental physics. How does this help us feel better about either physics or fairies? The truth is, if what one appreciated about physicalism, materialism or ontological reductionism was the beauty and simplicity of explanatory and ontological unity, strong emergence and panpsychism as forms of dualism are both gross disruptions to that picture of reality, just at different scales. Frankly, either view disconfirms the idea that matter traditionally understood and physics alone is fundamental. Substitute immortal souls for fairies and the point is clear, the mere act of putting conscious experience or subjectivity into fundamental physics doesn’t magically turn ``qualia'' or subjectivity into a physical property like momentum. Remember, the whole idea behind physicalism and materialism is to reduce or identify mental properties with biological or physical ones, not give them equal billing. For a detailed critique of panpsychism and dual-aspect theories see \cite{SilberSmolin,NMEntropy2020}.

What of consciousness and the hard problem? Given neutral monism, qualia is no longer the right way to conceive of conscious experience. While there are no doubt many important neural and information-theoretic correlates of conscious experience, with neutral monism, we do not tell a story about how fundamental quantum or neural processes dynamically or causally give rise to full-blooded conscious experience (i.e., panpsychism or strong emergence). We tell a story that starts with the nonduality of the so-called mental and physical (neutral monism), and then explain why we incorrectly perceive them as essentially distinct. How does that story go? Think of Kant’s unity of apperception: a minimal subject in a world in space and time go hand-in-hand, two sides of the same coin. As James puts it, ``not subject, not object, but object-plus-subject is the minimum that can actually be. The subject-object distinction meanwhile is entirely different from that between mind and matter, from that between body and soul. Souls were detachable, had separate destinies; things could happen to them'' \cite[p. 535]{james1904b}.

However, unlike Kant’s a priori transcendental condition (i.e., categories) for this ``object-plus-subject'' experience, neutral monism is proposing an a posteriori transcendental condition, not some cognitive or neural lens through which the noumenal is filtered. Neutral monism is a kind of direct realism. As James says, ``As `subjective' we say that the experience represents; as `objective' it is represented'' \cite[p. 480]{james1904a}. Indeed, fundamentally speaking, there is only ``the instant field of the present ... . It is only virtually or potentially either object or subject as yet. For the time being, it is plain, unqualified actuality or existence, a simple that'' \cite[p. 482]{james1904a}. What then is the self on this view? The self is simply subjectivity, a conscious PO to use the language from the beginning of the paper. As James puts it, ``The individualized self, which I believe to be the only thing properly called self, is a part of the content of the world experienced. The world experienced comes at all times with our body as its centre, centre of vision, centre of action, centre of interest. Where the body is is `here', where the body acts is `now'; what the body touches is `this'; all other things are `there', and `then' and `that' '' \cite{jamesPluralistic}. 

What then is the source of the illusion that self or subject is essentially distinct from the ``physical,'' ``external'' world? James says that the reification of subject/self only arises when, ``a given `bit' is abstracted from the flow of experience and retrospectively considered in the context of different relations, relations that are external to the experience taken singly but internal to the general flow of experience taken as a whole'' \cite[p. 535]{james1904b}. It is only the discursive intellect, in an inductive act of interpretation, that later creates or projects these dualisms between subject and object, inner/outer, self/world, etc. Keep in mind this discursive intellect is not some a priori cognitive category through which noumena is filtered. The mind’s inference to the dualism of knower/known, subject/object, etc., happens after the fact. And again, this is direct realism, there is no noumena. 

Given neutral monism, the mind and the world are one, just as Kant suspected. For Kant, given his unity of apperception, time is an a priori condition for experience, no subjectivity means no time or space. Kant here is providing a transcendental analysis in mentalistic terms. This means that the dynamical character of thought/experience and the world are two sides of the same coin. James puts it like this, ``According to radical empiricism, experience as a whole wears the form of a process in time''  \cite[p. 540]{james1904b}. Kant’s transcendental arguments from The Critique of Pure Reason are supposed to show that we must conceive of the world in a certain way, structure it internally according to certain categories such as time, space, and causation. Those arguments are fraught with many interpretative perils and controversies, but the basic idea is that experience is possible only if some experiences are conceptualized as being of enduring objects, enduring through time and space. Likewise, to experience a world of enduring objects there must be some sense of an enduring self. You cannot have one without the other. Again, however, as James notes, Kant was wrong that the structure of experience is a product or projection of mental filters or categories. Neutral monism takes the world of experience out of the head and also rejects the very idea of noumena or as some people call it, beables. Kant is right however that neither subject nor object alone, but only subject-object is the basic unit of experience. Nothing is mind dependent on this view in the subjective idealist sense. All entities and their properties are extrinsic or interdependent (not mind dependent!), not just colors, tastes and sounds, but mass, charge and spin as well. In short, mind and world are just two interdependent sides of the same coin. You can't have one without the other.

It should be clear that given neutral monism physics is not about the pursuit of some noumenal world or hidden magical beables, it is about the world of experience. It is common to make a distinction between metaphysical things-in-themselves and mere appearances or observations. Neutral monism rejects the very idea of the former, but the alternative is not anti-realism or sense data theory, it's radical empiricism. The idea that realism demands noumena or beables hiding behind the world of observables is simply a question begging misnomer. For us there is no inaccessible, noumenal QM realm hiding behind the world of experience. This is not to deny that under certain conditions, in certain contexts, reality behaves in ways that are best described by QM, e.g., as QM particles, fields or waves in spacetime. This is very much in keeping with the kind of thinking that led Einstein to relativity. As Jim Baggott puts it, ``In developing his theory of relativity, Einstein sought to banish from physics the entirely metaphysical concepts of absolute space and time. One consequence is that the observer is put firmly back into the reality that is being observed'' \cite[p. 109]{baggott2020}. Notice that there is nothing inherently anti-realist or instrumentalist in such a move. After all, Einstein is often described as the realist to Bohr's instrumentalist. We are not advocating for a ``shifty split'' as Bell put it, no magical line between the QM and the classical. However, for us, as evidenced by environmental decoherence, the QM and the classical are co-fundamental and co-dependent. This might violate a certain kind of fundamentalism or reductionism, but it doesn't violate commonsense realism.   

Just as the work of David Hume, Immanuel Kant, and Ernst Mach was essential for Einstein’s epiphanies behind SR \cite[p. 82-83]{isaacson}, so for us neutral monism (i.e., radical empiricism) not only shows us how to properly situate subjective experience in the world formally and metaphysically, but it shows how to move forward in physics as well. Indeed, in many ways these turn out to be the same project. Here is what Einstein said he gleaned from Hume and Mach, one must eliminate concepts that ``have no link with experience such as absolute simultaneity and absolute speed'' \cite[p. 131]{isaacson}. As James noted every scientific theory and all scientific explanations are, however opaquely, rooted in some metaphysical picture of the world such that ``the juices of metaphysical assumptions leak in at every joint'' \cite[p. 112]{taylor}. The radical empiricism (neutral monism) of James is just an extension of the empiricism of Hume, Einstein and others. In the words of Eugene Taylor \cite[p. 130]{taylor}:
\begin{quote}
Radical empiricism was, nevertheless, psychological; that is to say, it placed immediate experience at the center of everything we have to say about the universe. Consciousness, therefore, knower-and-known, subject-and-object, person-and-world, formed the basis of all science and all knowledge-getting. Positivistic science had to conform as much to the dictates of such psychology, as psychology was trying to conform to such a science.
\end{quote}
Radical empiricism is thus a metaphysics and epistemology of science. Taking it on board allows us to reconceive scientific explanation just as Einstein did with his principle explanation in relativity. This is precisely what we have done with QM as well. 

In Einstein’s words, ``the totality of our sense experiences ... can be put in order'' \cite{einstein}. We then use this model to explore regularities and patterns in the events we perceive. We mathematically describe these regularities and patterns and explore the consequences (experiments). In Einstein’s words again, ``operations with concepts, and the creation and use of definite functional relations between them, and the coordination of sense experiences to these concepts'' \cite{einstein}. We then refine our model of physical reality as necessary to conform to our results. This allows us to explain the past, manipulate physical reality in the present (to create new technology, for example), and to predict the future. While defining physics all the way down to individual ``sense experiences'' may seem unnecessarily detailed, it is crucial to understanding the relationship between subjective experience and physics being proposed here. In turn, we think that understanding will help the reader see how it could be the case that often the best explanations in physics are principle explanations, or explanations in terms of adynamical global constraints on sense experiences.

Putting this all together again, we can say the following. From our take on neutral monism we understand that each subject is just a conscious point of origin (PO) of Neutral Pure Presence. The perceptions of each PO form a context of interacting trans-temporal (enduring) objects (TTOs) for that PO. Since TTOs are ``bodily objects'' with worldlines in spacetime, TTOs are coextensive with space and time. When POs exchange information about their perceptions, they realize that some of their disparate perceptions fit self-consistently into a single spacetime model with different reference frames for each PO. Thus, physicists’ spacetime model of the ``Physical'' represents the self-consistent collection of shared perceptual information between POs, e.g., perceptions upon which Galilean or Lorentz transformations can be performed. Here is how Hermann Weyl himself put it, Physics is the ``Construction of objective reality out of the material of immediate experience'' \cite[p. 117]{ryckman}.

Coordinate transformation is important because relativity states that there is not one reference point (or perspective) in the universe that is more favored than another. As Weyl put it, ``The explanation of the law of gravitation thus lies in the fact that we are dealing with a world surveyed from within'' \cite[p. 117]{ryckman}. Keep in mind that the beauty of neutral monism is that talk about POs and their perceptions should be understood not as some sort of positivism, or some brand of idealism (subjective or otherwise), or  sense data theory, or merely as bracketed phenomenology, but in terms of James' ``instant field of the present'' and what Russell calls ``events.'' That and that alone is what spacetime is. And spacetime is the subject of physics.

This brings us to our axioms that we stated at the beginning of our paper. Given neutral monism, it should be clear why we chose those particular principles as the basis for all of physics. That is, the universe (as experienced) is the self-consistent collection of shared classical information regarding diachronic entities (classical objects), which interact per QM. The consistency of shared classical information of the  universe is guaranteed by the divergence-free (gauge invariant) nature of the adynamical global constraints for classical and quantum physics. In the case of Wigner's friend per Healey, the self-consistent collection of classical information would include all shared classical information between Xena, Yvonne, Zeus, and Wigner. In the case of the delayed choice quantum eraser experiment, conscious choices are equally constrained. Here we see a profound connection between QM, relativity, and conscious experience. But, we do not think this is any weirder than the fact that conscious experiences and choices are constrained by other adynamical global constraints, such as conservation laws, the light postulate, and the relativity principle. Due to limits of time and space, we cannot recapitulate the work along these lines, but for those interested, we showed how to derive QM and relativity from our axioms \cite{NMEntropy2020}, which in turn gives us the tools to address the experiments herein.

Let us recall our goal for this paper was to provide a take on QM that explains why there is and must always be determinate and intersubjectively consistent experience about all experimental outcomes (absoluteness of observed events). A take that accepts the completeness of the theory and requires no invocation of relative states (e.g., outcomes being relative to branches, conscious observers, etc.). And finally, a take that requires no allegedly hybrid models such as claims about ``subjective collapse.'' We wanted a take on QM that yields a single world wherein all the observers (conscious or otherwise) agree about determinate and definite outcomes, because those outcomes are in fact determinate and definite. We wanted a realist psi-epistemic take on QM that saves the absoluteness of observed events and the completeness of QM, without giving up free will or locality. We also wanted to show how our realist psi-epistemic account eliminates the measurement problem and, coupled with our take on neutral monism, also eliminates the hard problem of consciousness. We believe we have done all of the above. The key to achieving these goals was to let go of the following offending assumptions: 1) physicalism, 2) fundamentalism, and relatedly 3) dualism about conscious experience, 4) the notion that fundamental explanation is always constructive, causal or dynamical, and relatedly, 5) realism about the wavefunction. Together these assumptions force us into the hard problem, they force us into the measurement problem, and they force us to seek the solutions to these problems in fundamental physics, e.g., by trying to relate these problems to one another directly, with very little success. Once again, sometimes, when a problem is deeply intractable the best move is to jettison the offending assumptions that led to the problem in the first place. This is precisely what we did herein.

\bibliography{biblio.bib}

\end{document}